\documentstyle[draftjgr,psfig]{article}
\begin{document}

\title{\bf Earthquake source parameters and fault kinematics
in the Eastern California Shear Zone} 
\vspace{0.25in}
\author{Laura E. Jones and Donald V. Helmberger}
\address{Seismological Laboratory, California Institute of
Technology, Pasadena, California}
\maketitle
\vspace{0.25in}

\section{Abstract}

Based on waveform data from a profile of aftershocks following
the north-south trace of the June 28, 1992 Landers rupture across
the Mojave desert, we construct a new
velocity model for the
Mojave region which features a thin, slow crust.
Using this model, we obtain source parameters, including depth and
duration, for each of the aftershocks in the profile,
and in addition, any significant ($M>3.7$) Joshua Tree--Landers
aftershock between April, 1992 and October, 1994
for which coherent TERRAscope data were available.
In all, we determine source parameters and stress-drops for 45
significant ($M_w > 4$) earthquakes associated with the
Joshua Tree and Landers sequences,
using a waveform
grid--search algorithm.
Stress drops for these earthquakes appear to vary systematically
with location,  with respect to previous seismic activity,
proximity to previous rupture (i.e., with respect to the
Landers rupture), and with tectonic province. In general,
for areas north of the Pinto Mountain fault,
stress-drops of aftershocks located off the faults involved
with the Landers rupture are higher than those
located on the fault, with the exception of aftershocks on
the newly recognized Kickapoo (Landers) fault.
Stress drops are moderate
south of the Pinto Mountain fault, where there is a history
of seismic swarms but no single through-going fault.
In contrast to aftershocks in the eastern Transverse
ranges, and related to the 1992 Big Bear, California, sequence, Landers
events show no clear relationship between stress--drop and
depth. Instead, higher stress--drop aftershocks appear to correlate
with activity on nascent faults, or those which experienced
relatively small slip during mainshock rupture.

\section{Introduction}

Stress-drop and style, depth and timing of aftershock activity 
relative to mainshock  rupture plane or fault trace yields clues
about how the regional `stress budget' is settled following  a
large earthquake.   Aftershock stress-drops vary  with 
source area and tectonic environment [Lindley and Archuleta, 1992],
reflecting regional differences in the source
properties of small earthquakes. 

The $M_w7.3$ Landers earthquake of 11:58 GMT, June 28, 1992, was
preceded by the April 23, 1992, Joshua Tree mainshock ($M_w6.1$)
which is now considered
a precursory event [Stein et al., 1994] with its own
substantial fore-- and aftershock sequence.
The Landers event was followed
by tens of thousands of aftershocks [Kanamori et al.,
1992; Hauksson et al., 1993; Sieh et al., 1993], many in
areas with no surface rupture [e.g, Big Bear region,
see Figure 1].
Stress-drops and source parameters of Joshua Tree--Landers
aftershocks provide information critical to understanding fault
kinematics in the Eastern California Shear Zone (ECSZ), which
encompasses the Landers rupture area and may extend beneath
the eastern Transverse ranges [Jones and Hough, 1995].

Because data for the present study comes from a sparse array
[three to five TERRAscope stations], care must be taken
when modeling available data to ensure accuracy in depth and
source mechanism estimation.  A standard one-dimensional
model such as the Southern California Model may often be used to
satisfactorily approximate broadband waveforms at near-regional
distances [see Dreger and Helmberger, 1991]. However,  waveform
misfit introduced by use of an inappropriately thick crust,
for example, more adversely affects quality and robustness
(error) of source solutions obtained from small datasets.
A regional model is thus necessary for this work.

In this paper, we present source parameters, including duration,
depth, and stress-drop, obtained for Landers and Joshua Tree events
using a new earth model designed to fit near-regional data with  source--
receiver paths in the Mojave. The paper treats events from this
large sequence as follows, moving chronologically from the April,
1992 Joshua Tree `pre-shock', to Landers aftershocks, first south
and then north of the Pinto Mountain Fault, including a cluster of
events in the Barstow region and triggered quakes on the Garlock
fault. Events occuring within the ECSZ are compared with
Landers aftershocks occuring in the Eastern Transverse Ranges
and comprising the 1992 Big Bear, California, sequence.
Finally, we correlate aftershock stress-drop with timing
and proximity to mainshock rupture.

\section{Data and Observations}

Larger fore-- and aftershocks from
the Joshua Tree and Landers sequences were recorded
on scale by six broadband TERRAsope stations (GSC, ISA, PAS, PFO,
SVD and SBC). In this study we use records from the first five
stations [Figure 2], since records from station SBC are low signal-to-noise, and
contaminated by propagation through basin structure.
For TERRAscope stations Goldstone (GSC) and Pinyon Flats (PFO),
due north and nearly south of the Landers rupture, we
construct profiles of aftershocks from the Landers earthquake.
These include earthquakes in areas associated with Landers surface
rupture (north of the Pinto Mountain fault), south of the
Pinto Mountain fault, and associated with the Barstow swarm.
These earthquakes form rough profiles following the general
trend of the Landers rupture.

Before modeling, the records were processed as follows:
instrument gain was removed from the raw velocity records;
they were detrended and integrated once. A butterworth
bandpass filter with corners at 0.04 and 7 Hz was applied
twice. Filtering was minimal so that the broadband nature
of the records might be preserved. In cases where the
event was fairly large and close to a particular station,
low--gain records (accelerograms) from TERRAscope were used.
They were processed similarly: gain removed, detrended, twice
integrated, and bandpass filtered.

\section{Analysis}

\subsection{The Mojave Model}
Studies to date on moderately--sized Southern
California earthquakes suggest that a relatively simple,
plane--layered velocity model often explains the
observed waveforms satisfactorily. For example,
waveforms from the June 28, 1991 Sierra Madre
earthquake, centered within the TERRAscope array,
were well--modeled at several stations 
by the Standard Southern California model
[Hadley and Kanamori, 1977; Dreger and Helmberger, 1991].
Studies of several other events also suggest
that this standard model is appropriate for use in the
Southern California region [Jones and Helmberger, 1995;
Song and Helmberger, 1997]. However, this
standard model did not work well for
Landers aftershocks recorded at stations in the Mojave Desert.

High--quality aftershock data recorded at
local to regional distances
gave us the opportunity to develop a
path--specific model for the Mojave region.
Aftershocks from
the Landers sequence recorded at TERRAScope stations
Goldstone (GSC) and  Pinon Flats (PFO) were assembled,
and profiles of broadband data constructed from events
located and recorded in the Mojave block, as such possessing
source--receiver paths contained entirely within
this region [Figure 2].
Records at these distances (35-165 km, see Figure 3)
are dominated by crustal arrivals and Moho--reflected arrivals,
which suggest a crust thinner (depth to the Moho is 28 km)
and slower than the standard Southern California Model
[Hadley and Kanamori, 1978; Dreger and Helmberger, 1991]
and lacking the gradient at the base of the
crust (Conrad) which characterizes the widely
used Standard Model.

The choice of stations GSC and PFO for this modeling task was natural and
fortunate, since Landers events recorded at these two
stations form north--south profiles.
The locations of stations
GSC and PFO nearly due north and south (respectively) of the
aftershocks, however, practically insures that many events will be
{\sl P}--wave nodal at both stations, since many
have northerly strikes (parallel to the Landers
rupture). 
Conversely, the tangential component is at or near
maximum, so it is easily modeled [Figure 3]. 
 
\begin{table}
\begin{center}
\centerline{Table I: Standard Southern California Model}
\small
\begin{tabular}{ccccc}
\hline
   $V_p$  & $V_s$ & $\rho$  & depth \\
   (km/s) & (km/s) & (g/$cm^3$) & km & \\
   \hline
     5.50 &    3.18 &     2.40   &     5.5  \\
     6.30 &    3.64 &     2.67   &    16.0  \\
     6.70 &    3.87 &     2.80    &   32.0  \\
     7.85 &    4.50 &     3.42   & half space \\
\hline
\end{tabular}
\end{center}
\label{lantab1a}
\end{table}

In order to construct
the model, we first make an estimate of the source
mechanisms for the profile events, assuming the
standard Southern California model [Table I].
We subsequently refine the original source and
moment estimations for the profile events using the new model;
these estimations show improved
waveform fit, and lower error.
 
The Mojave model[Table II] has a thinner crust (28 km versus 35 km) than the
standard California model, and slower {\em P} and {\em S} wave
crustal velocities. It also lacks the gradient at the
base of the crust (the so--called ``Conrad'' discontinuity)
which characterizes the standard model.

\begin{table}
\begin{center}
\centerline{Table II: Mojave Model}
\small
\begin{tabular}{ccccc}
\hline
    $V_p$  & $V_s$ & $\rho$  & depth \\
    (km/s) & (km/s) & (g/$cm^3$) & km & \\
   \hline
   5.00 &    2.60 &     2.40   &     2.5   \\
   5.50 &    3.45 &     2.40   &     5.5  \\
   6.30 &    3.60 &     2.67   &    28.0  \\
   7.85 &    4.40 &     3.42   & half space \\
\hline
\end{tabular}
\end{center}
\label{lantab1}
\end{table}
 
\subsection{Determination of Source Parameters}
Average source parameters and depths for the small and
moderately sized earthquakes studied here are
 estimated using a direct grid--search method
[Zhao and Helmberger, 1994]. This algorithm
selects source parameters which minimize the L1 and L2
norms between observations and synthetic waveforms, using
three component  $P_{nl}$ 
and whole waveforms 
to produce a stable solution from
a relatively sparse data set and an imperfect structural
model [Jones et al, 1993; Jones and Helmberger, 1995; Zhu and
Helmberger, 1996; Song and Helmerger, 1996]. 
Note that $P_{nl}$ is defined as the first part of the regional
waveform, from where the record is dominated by $P$ phases ($P_{n}$) to
where the motion contains progressively more $SV$ contributions ($PL$)
[Helmberger and Engen, 1980].
The procedure desensitizes the misfit in timing between principal
crustal arrivals in the data and synthetic by fitting portions
of the waveforms independently.
Source durations for the grid-search are initially estimated from
the width of the direct pulse. Refined durations (see below) are
then iteratively fed back into the grid search scheme to recompute
source parameters.
Given the development of Green's functions specific to paths
within the Mojave block, we use a sparse array (three to five stations)
and the data both
broadband and after convolution with a long--period
Press-Ewing (``LP3090'': 30 s period, 90 s galvenometer)
instrument response.  The long--period
energy is modeled because the solutions are often more stable
than the broadband solutions, as detailed below,
though we seek consistency 
between broadband and long--period solutions.
Broadband solutions were occasionally used for the smallest
events, in cases where energy was lacking in the long-period
bandpass and the broadband solution showed greater consistency
between stations.

\subsubsection{Estimation of Source Depths}\\
We determine source depths directly from the surface
reflected phases $S_mS$ or $sS_mS$, and by cycling through
depth--dependent Green's functions (2, 5, 8, 11, 14, and 17 km)
during the grid-search procedure itself.
To speed the process we employ a catalog of Green's functions
appropriate to the Mojave
model, which are computed at 5 km distance intervals from 35 km to
400 km, and assuming source depths listed above. 
In general, the mechanisms and depths obtained in this
study are consistent with those obtained by other workers.
In some cases, however, the
depths we obtain are not as shallow as those obtained by
others [Thio, 1996, by surface wave inversion;  Hauksson, 1993,
via inversion of short--period network data]. 
As an example, we show modeling for the August 5, 1992, 22:22
GMT Landers aftershock (Figure 4).
Fits for all three
components (including the radial) are shown. Error space for
the depth determination (Figure 5) shows a clear minimum at
between 5 and 8 km for this event, though others place the depth
of this event at less than 5 km [Hauksson, 1993].
$P_{nl}$ to surface--wave amplitude ratios on the vertical and radial
components of motion suggest a depth of about
5 km, while ratios of body wave to Love wave amplitudes
suggest a depth of 8 km or greater.
Indeed, separation between $S_mS$ and $sS_mS$ phases on the
tangential components at stations PFO (epicentral distance
155 km), ISA (160 km) and PAS suggest a depth arguably deeper
than 8 km.

Within the error imposed by the depth gridding on
our solution space (every 2-3 km), we believe that our depths,
obtained from a grid--search routine which is tantamount to
direct waveform modeling, are reliable. There is substantial
difference in the separation between $S_mS$ and $sS_mS$ phases
for events at source--depths of, say, 2 and 5 km. Our estimates
suggest that all of the $M>3.7$ events we studied had depths of
5 km or greater; and average depth is about 8 km.

\subsection{Source Duration and relative Stress-Drop}
Source durations are obtained by methods ranging from
direct measurement of source pulse [e.g., Smith and Priestly,
1993;  Hardebeck
and Haukssen, 1997], to determination of corner frequency
[e.g., Hough and Dreger, 1995].
In this study, average source durations are determined 
from a simple comparison of energies [see also
Jones and Helmberger, 1996; Zhao and Helmberger, 1996;
Song and Helmberger, 1997].
In this procedure, we equalize energy content across
different frequency bands between data and synthetics.
First, a short-period Wood-Anderson
instrument response (WASP) and a long--period instrument
response (LP3090) are applied to data and synthetics
to compute short-- and  long--period energy, respectively.
The $P_{nl}$ waves (in velocity) from each station are
then compared with synthetic
$P_{nl}$ waveforms (velocity):
 
\begin{equation}
Ratio = \frac{E_{(obs)}}{E_{(syn)}}
\end{equation}
 
\noindent
where
 
\begin{equation}
E = \frac{\int_{t_{pn}}^{t_{PL}}{[V_{(sp)}]^{2}}dt}
{\int_{t_{pn}}^{t_{PL}}{[V_{(lp)}]^{2}}dt}
\end{equation}
 
\noindent
$V_{(sp)}$  is the observed (or synthetic)
$P_{nl}$ wave, in velocity, convolved with
a short-period Wood--Anderson response, while $V_{(lp)}$ is the
observed (or synthetic) $P_{nl}$ wave, in velocity, convolved with
an LP3090 instrument response. The time-function for the synthetic
waveform is adjusted until the ratio of energies is
unity (symmetric trapezoidal time functions are assumed).
An average for the radial and vertical components is 
found at each station, and the resulting values for each reporting
station are then averaged.

The procedure yields a conservative
estimate of source--time duration and thus stress-drop,
and is limited to source
triangles no shorter than $0.20$ s in duration. This limitation
is imposed by the computational technique used, and to a lesser
extent, by the frequency content available in the synthetic Green's
functions. Other researchers using this method 
found good correlation between source durations determined via
comparison of energies and those determined by measuring the width of
the direct pulse at local stations [ Song and Helmberger, 1997],
except for a (constant) offset. The offset may be explained 
by the fact that the synthetics used in the energy method do not contain
scattering [Song and Helmberger, 1997]. Note that source durations obtained by
energy comparison are systematically smaller
than those obtained via direct measurement. The energy method thus
provides a reliable estimate of `relative' source duration between
events.
 
Assuming minimal attenuation, the width of the observed
{\em P} or {\em S} pulse is 
proportional to the source dimension, and thus source duration.
The actual pulse-width, as observed, may depend on factors as
diverse as crustal attenuation, rupture mode, length and velocity,
and source complexity. On average, however, it is acceptable
to assume a linear relationship between pulse--width and
source dimension.
Indeed, Cohn et al. [1982],
assuming a circular fault [Brune, 1970],
obtained the relation

\begin{equation}
\tau=\frac{2.62a}{\beta}
\end{equation}

\noindent
where $\tau$ is the source duration in seconds, $a$ is the
radius in km, and $\beta$ is the shear velocity local to the
source region. Solving for $a$ in terms of $\tau$, assuming a
shear velocity of $3.5$ km/s, and substituting the result
into the expression for stress--drop on a circular fault
[Eschelby, 1957]

\begin{equation}
\triangle\sigma =\frac{7M_{o}}{16a^{3}}
\end{equation}
 
\noindent
we obtain (in bars, given 1 bar = $10^6$ dyne-$cm^2$)
 
\begin{equation}
\triangle\sigma=\frac{1.84\times10^{-22}M_{o}}{\tau^{3}}
\end{equation}
 
\noindent
An estimate of the error inherent in the computation of
relative stress-drop is found as follows. Assuming that the
error in $M_{o}$ and $\tau$ are to first order independent,
we can write the error as the vector sum of error in $\triangle\sigma$
due to  error in the estimates of $M_{o}$ and $\tau$, respectively:

\begin{equation}
\delta[\triangle\sigma] = \sqrt{(\frac{\delta[\triangle\sigma]}{\delta\tau}\triangle\tau)^2 + (\frac{\delta[\triangle\sigma]}{\delta M_{o}}\triangle M_{o})^2}
\end{equation}

\noindent
Taking partial derivatives of (5) with respect to $\tau$ (holding $M_{o}$ constant) and
$M_{o}$ (holding $\tau$ constant), substituting into (6) and simplifying,

\begin{equation}
\delta[\triangle\sigma] =  \sqrt{[\frac{3\triangle\sigma}{\tau}\triangle\tau]^2 + [\frac{\triangle\sigma}{M_{o}}\triangle M_{o}]^2}
\end{equation}

\noindent
Factoring out a $\triangle\sigma$ in (7), we obtain percentage error:

\begin{equation}
\frac{\delta[\triangle\sigma]}{\triangle\sigma} =  \sqrt{[\frac{3}{\tau}\triangle\tau]^2 + [\frac{1}{M_{o}}\triangle M_{o}]^2}
\end{equation}

\noindent

Small events with shorter time functions had
relatively greater error associated with the determination of
source duration, and often greater error associated with the
determination of moment (due to poor signal to noise).
 For the Joshua Tree sequence, for example,
 we obtain errors ranging from 
67\%, for an event
with 58\% error in the moment estimation and $M_b 4.3$, to
$32\%$,  for an event with 29\% error in moment estimation,
and $M_b 4.5$. Larger events are predictably associated with smaller
error. The July 11, 1992, $M_b 5.1$  Garlock fault
event had an uncertainty in moment estimation of 24\%,
and an error in stress-drop estimation of about 20\%.

We use relative stress-drop along with source parameters
in the following discussion to explore the relation
between source type, depth, location and relative energy
release in the eastern California shear zone.

\section{Results and Discussion}
Large aftershocks occurring up to two-and-a-half
Coulomb stress changes
caused by four $M>5$ earthquakes preceding the Landers
mainshock (i.e., the 1975 $M_L5.2$ Galway Lake, 1979
$M_L5.2$ Homestead Valley, $M_L6$ North Palm Springs
and $M_L6.1$ Joshua Tree earthquakes) progressively
increased stresses at the site of the future Landers
epicenter [King et al, 1994]. In turn, changes in
static stresses caused by the Landers event triggered
the Big Bear event within hours of the Landers mainshock,
and earthquakes as far away as the western Garlock fault
and Yucca Mountain in the ensuing months
[Hill et al., 1993, Gomberg and Bodin, 1994].

As discussed below, Joshua Tree sequence seismicity moved northwards
in the months following the Joshua tree mainshock, 
culminating in clusters of aftershocks just north of the 
Pinto Mountain fault and within the Landers epicentral
area in early June of 1992.
Hours before the Landers mainshock, a 
cluster formed at what later became the Landers epicenter
\cite{hauksson}. The Landers earthquake involved rupture
on five separate faults north of the Pinto Mountain fault,
with a small amount of displacement south of the Pinto
Mountain fault on the Eureka Peak fault (Figure 6).
The latter rupture may not have occurred entirely during the
mainshock, but may have been associated with a $M5.7$
aftershock occurring minutes after the mainshock \cite{Hough}.
 
We divide our discussion of the Landers sequence
into four portions: aftershocks south of the Pinto Mountain
fault, including the Joshua Tree 'preshock' sequence, and
associated with minimal displacement; aftershocks
north of the Pinto Mountain fault, associated with the
Landers rupture, aftershocks north and east of the mapped
Landers rupture, in the Barstow and  Calico--Pisgah fault
clusters, respectively; and aftershocks or triggered events
along the Garlock fault.

\subsection{Joshua Tree Sequence}
The Joshua Tree sequence
began on April 23, 1992 at 02:25 GMT
with a $M_w=4.3$  foreshock. This event occurred at
a location just south of the Pinto Mountain fault
(-116.32 W, 33.94 N),
and north of the Coachella Valley segment of the San Andreas
fault, within the Little San Bernardino Mountains, in a region
which has historically seen frequent earthquake swarms.
It was followed by a number of additional smaller foreshocks, then
within two-and-a half hours by the nearly co-located
$M_w=6.1$ Joshua Tree mainshock (Mori, 1994).
The Joshua Tree mainshock
had no observed surface rupture, though a 10-to 12 km south-to-north
subsurface fault--plane, striking roughly $N20^oW$,
was inferred from the distribution of early aftershocks
[Wald, personal comm., 1992; Hauksson et al., 1993;
Hough and Dreger, 1994].
 
The mainshock was
followed by a sustained and powerful aftershock series which
comprised at least 28 aftershocks of $M>3.7$, 10 of which were
$M4.0-M4.7$. 
Joshua Tree aftershocks partially overlap
those from the later Landers earthquake, with a cluster of
aftershocks, including one event above $M4$, developing north of the Pinto
Mountain fault and slightly east of the Landers mainshock location
in early June (e.g., Figure 7a, aftershock number 9).
$M>3.9$ aftershocks form two separate clusters south of the
Pinto Mountain fault which are filled in by later
aftershocks from the Landers earthquake (Figure 7b).
The Joshua Tree series is dominated by moderate to deep
(source depth 8--14 km) strike--slip and oblique--slip
events.
Stress-drops for
these earthquakes are on the order of
$10-100$ bars, with an average of 30 bars. 
 
Events of the Joshua Tree sequence are
now viewed as preshocks to the later Landers mainshock. While
the Landers mainshock
apparently either recharged or ``reactivated'' aftershock activity
in the Joshua Tree region \cite{haukssonB}, $M>3.8$ aftershocks
from the Joshua Tree and later Landers events can be viewed as
distinct populations. Spatially, they occupy distinct but
ajoining volumes
rather than overlapping completely
(Figures 7b, 8).  Their
mechanisms are similar, presumably strike--slip on north to
northwest--striking planes, though Joshua Tree aftershocks are
on average deeper [Tables III, IV].
The presence
of several $M>4$ Landers aftershocks in the Joshua Tree epicentral
region supports post--Landers reactivation of stresses
immediately local to the Joshua Tree epicentral area. 
These $M>4$ events are not numerous, however, are low in
stress--drop relative to other aftershocks south of the
Pinto Mountain fault, and are generally not vertical strike--slip.

\subsection{Landers events south of the Pinto Mountain Fault}
Following the Landers mainshock,
large ($M>4.5$) aftershocks were more common
{\em south} of the Pinto mountain fault than
north (Figures 8, 9, 12).
Almost 76\% of the total aftershock energy released
post--Landers was released south of the mainshock epicenter,
with about 40\%  of the energy release distributed between
the Pinto Mountain fault and the Joshua Tree epicenter
\cite{ma}.
 
A tight and dense cluster of early aftershocks formed near the
epicentral locations of the events on the
Eureka Peak and Burnt Mountain faults, as observed in
the immediate aftermath and epicentral location of the
(northern) Landers mainshock (Figure 8). Unlike the Landers
epicentral area, however, large ($M\ge4$) aftershocks continued
in this southern region for many months.

Aftershocks extend roughly 40 km south
of the mainshock epicenter, forming a NW--SE trending
swath 5-15 km in width \cite{hauksson}. We present
source parameters, depths, durations and relative stress-drops
for 14 $M_w \geq 3.7$ aftershocks occurring south of
the Pinto Mountain fault, including an $M_w4.5$ event
on August 21, 1993
(Figure 9, Table IV, event number 13)
and two events in August of 1994
(Figure 9, Table IV, events 14--15).
Events studied suggest a fairly heterogeneous sequence,
though oblique strike--slip events are most numerous.
These oblique events are consistent in strike direction;
all strike NW, presumably in the same direction as the Joshua Tree
mainshock ($N20^oW$) and with strike--slip
events associated with the Joshua Tree sequence
(Figure 7ab).
 
Like those estimated for Joshua Tree aftershocks, relative
stress-drops for Landers aftershocks south of the Pinto Mountain
fault are on the order of $10-100$ bars; with an average of about 67
bars for aftershocks within the first year of the mainshock, and
an average of 60 bars
for aftershocks through 1994.
Lowest stress--drop events are associated with either the
epicentral region of the southern rupture (Figure 7b),
or the area active during
earlier Joshua Tree sequence (including the Joshua Tree
mainshock) located south of the southern rupture.
High stress--drop earthquakes (events 2, 9, 10) lie west and nearly on the
periphery of the low stress--drop cluster associated with Eureka Peak
rupture (i.e., events
3, 4, 5, 8, 12) as
seen in Figure 9.
Event 14  (on the periphery of former Joshua Tree seismicity)
is unusually low stress-drop, but occurred
after much of the sequence had exhausted itself:
this late $M_w3.7$ event occurred
in August of 1994, at a depth of 8 km.
On average, Landers events are higher stress-drop
than Joshua Tree
events [Tables III, IV, Figure 10], again supporting the
notion that the Landers
mainshock may have recharged this historically active region.

In map view 'Southern Landers' events do not define any one fault plane; rather
they re-rupture areas associated with the Joshua Tree sequence, and
fill in unaffected regions north towards the Pinto Mountain
fault. The history of seismic activity in the region, the present
heterogeneity of faulting and the
lack of any one well-defined fault plane suggest that displacement
south of the Pinto mountain fault
may be accommodated gradually (i.e., in small increments) across a number
of small subsurface faults.
The gap in large aftershocks across the Pinto Mountain fault
(Figure 9) suggests that Landers rupture
may not continue across
the fault, and that displacement south of the Pinto Mountain fault
may be primarily associated with aftershock activity.
 
\subsection{Landers events occurring North of the Pinto Mountain Fault}
Rupture along the five faults active in the Landers
mainshock (from south to north, the Johnson Valley fault,
the Kickapoo (Landers) fault, the Homestead Valley fault,
the Emerson fault and the Camp Rock fault) extended roughly
60 km N-NW across the Mojave desert north of the Pinto Mountain
fault (Figure 6).
Large ($M>3.9$) aftershocks along the trend of the Landers
rupture are common in three general areas:
close to the mainshock epicenter (early aftershocks, within
the first 24-48 hours), at fault ends, including the
termination of the Johnson Valley fault and the very active
Kickapoo (Landers) fault, and the northern extent of
rupture, at the northern terminus of the Camp Rock fault
(Figures 6, 12). Landers aftershocks north of the Pinto
Mountain fault (discounting events on the Garlock) are
higher stress-drop than southern Landers aftershocks, with 
on average of 95 bars for events occurring in the first
year after the mainshock [Figure 10, Table V]. 

\subsubsection{Mainshock Epicentral Area (Johnson Valley Fault)}
According to Wald and Heaton [1994], the Landers mainshock
initiated on the Johnson Valley fault (JVF) at depth, and the
first seconds of rupture involved deep slip. Rupture
then continued shallowly on the JVF for the subsequent 4 seconds.
The region immediately local to the Landers epicenter, along
the previously recognized and active Johnson Valley fault, saw
many $M>4$ aftershocks within the first 24 hours of the
mainshock,~\cite{hauksson}. However, we were not
able to obtain TERRAscope data for these early events. We
examined two later events, one nearly co--located with
the mainshock (Figure 12, event 10), and one slightly northeast
of the same, a $M4.7$ event which occurred in June of 1994
(Figure 12, event 19). Both events are oblique--slip, 
of moderate to shallow source depth, and are low stress-drop
(9 and 15 bars, respectively,
see Figures 12 and 13), suggesting that
stresses local to the mainshock epicentral area were fairly
low in the months and hours following the Landers earthquake.
Indeed, according to Abercrombie and Mori [1994], the mainshock
itself began with a shallow, low stress-drop 
preshock composed of two $M\sim4-5$ subevents
(stress--drops for both $\sim12$ bars),
which triggered or grew into the $M7.3$ Landers mainshock.

\subsubsection{Kickapoo (Landers) Fault}
There were an unusual number of $M_w>3.9$ aftershocks along
the short segment of the newly recognized Kickapoo (Landers)
fault. This is a previously unmapped, 5 km long N-S trending
fault strand running from the northern leg of the Johnson
Valley fault northwards to the southernmost end
of the Homestead Valley fault. Rupture during the 1992
Landers event propagated from the Johnson Valley fault to
the Homestead Valley fault along the Kickapoo fault and
secondary fault traces just east of the Kickapoo \cite{sowers}.
We studied four (out of six)
$M>3.9$ aftershocks occurring along or near the Kickapoo fault which
were recorded on the TERRAscope array [Figure 12].
 
The earliest event
is a normal-faulting event  occurring
near the southern end of the zone comprised of the Kickapoo
and its secondary faults (Figure 12, Table V, event 1). 
It is of moderate stress--drop (84 bars)
and average depth for this region.
It was followed by two strike--slip to oblique--slip events just
north along the Kickapoo (Figure 12, Table V, events 3 and 9)
The first of these is the largest
aftershock to occur within the Landers rupture region, at
$M_w=5.2$, and also has the highest stress-drop (about
$515\pm176$  bars, Table 
A second ($M_w3.9$) colocated right-lateral
strike-slip aftershock occurred two weeks later (event 9) at
a depth of about 6 km.
This event is substantially smaller, has a much lower stress-drop
(30 bars), and may represent re-rupturing of a previously
ruptured fault-patch. A later $M_w4.3$ event occurred near
the southern end of the Homestead Valley fault approximately
near the termination of the Kickapoo fault (Figure 12, event 15).
This aftershock is of similar depth (7 km), has an
oblique--slip source mechanism,
and a stress-drop of about 86 bars. It occurred within a region
mapped and described by Spotila and Sieh [1995], and exhibiting
both strike--slip and thrust faulting. This region was associated
with a slip-gap during the Landers rupture, and showed some vertical
offset but virtually no strike-slip motion.
 
The presence of the latter three events lends support to the  dominantly
right-lateral offset ``through-going''
model suggested by Sowers et al. [1994] for the Kickapoo fault.
However, the mechanism of the earliest large Kickapoo aftershock
(event 1) suggests extension, which lends credence to the less
favored ``step-over model'' suggested by Sowers et al. [1994].
Clearly the tectonics of the Kickapoo fault is more complicated
than either of these simple schemes; perhaps some combination of
the two models, might explain
the complex seismicity we observe here. The presence of so many
heterogeneous and high stress--drop aftershocks along this small
segment of fault also lends credence to the suggestion made
by Spotila and Sieh [1995], that the connection between the
Johnson Valley and Homestead Valley faults is incomplete,
and that the Kickapoo fault is still very immature.

\subsubsection{Emerson and Camp Rock Faults}
Large on--fault aftershocks appear to be much less common
north of the Kickapoo Fault. Most $M>3.9$ aftershock activity
appears to be concentrated near the end of rupture on the
Camp Rock fault. Relative stress--drops on these faults
are low to  moderate, ranging from 38 to 86  bars for the
events we studied [Table IV].

\subsection{Off--Fault Aftershock Activity}
In addition, there are clusters of large aftershocks
off-fault (i.e., unrelated to any primary rupture during
the Landers mainshock). These occurred east of the Landers
rupture, near the Pisgah/Calico faults (Figures 12 and 15)
and north of the terminus of Landers rupture on the
Camp Rock fault, in the Barstow region.

\subsubsection{Aftershocks on Pisgah--Calico Faults}
Aftershocks near the Calico fault [Figure 11] form two
east-west alignments perpendicular
to the trend of the Landers rupture, roughly at the latitudes
of the Emerson and Camp Rock faults [Figure 6].
Stress--drops
for two $M>4$  events [Figure 11, events 16, 20] are moderate to high; the
latter event (20) occurred more than two years after the Landers
mainshock but shows similar fault motion and depth as the
earlier event (16) occurring in August, 1992.
In addition, there is a spatially and temporally
tight cluster of aftershocks
just east of the Pisgah fault, several of which are larger than
$M4$.
Two of these occurred within an hour of each other,
and were nearly colocated (events 5 and 6); the second
event having a lower relative stress drop (25 bars)
than the first (71 bars).
Aftershocks on the Pisgah and Calico faults may be related
to off--fault strain caused by changes in strike along the
Landers rupture \cite{sieh}.  High stress--drops in both regions
might suggest high applied shear stresses along north to
northwest--striking planes.

\subsubsection{The Barstow Sequence}
The Barstow cluster was associated with no surface rupture,
and occurred approximately 30 to 40 km north of the aftershocks
associated with northernmost Landers rupture on the Camp Rock
fault. It began approximately 6 hours after the Landers mainshock,
and comprised at least 12 aftershocks above $M4$.
The largest aftershock,
at $M_w=4.4$, occurred on August 5, 1992, at 22:22 GMT, within a
tight cluster of larger aftershocks towards the southern end
of the trend [Figure 13]. The Barstow sequence is fairly narrow 
in width compared
with aftershocks along the Landers rupture; the ratio
of length (about 20 km) to  width (2--3 km) has been cited as
evidence that the Barstow sequence may have occurred on a single
fault, unlike Landers [Hauksson et al., 1993].
However, closer examination of the larger aftershocks in the
sequence shows a distinct jog in the trend of the
aftershocks, with a tight cluster to the southeast (e.g., aftershocks
11, 14, at depths of  8 and 7 km, respectively) which could arguably
have occurred on a single fault. There is an abrupt step-over, with
events farther to the west
(events 4, 11) along a rough
trend striking NW--SE. Stress--drops for these earthquakes range
from $16-80$ bars, with an average of about 50 bars.
Our depth estimations do not show the shallowing reported by
Hauksson et al.,[1993], and shallowest events are at a depth of
5 km.

\subsection{Aftershocks or 'Triggered Events' on the Garlock fault?}

The Garlock fault has long been recognized as an important
tectonic feature in Southern California.
Though it has not
produced any large earthquakes within the period of historical
record, numerous scarps and left-laterally offset Holocene
features suggest that the fault is active and has
produced large earthquakes. As recent levels of
seismic activity on this fault are low in comparison to those
inferred from Holocene displacements, the Garlock fault
may represent a seismic gap \cite{astiz}. Until the moderately sized
earthquakes in July of 1992 [Figure 13, event 8]
and again in October, 1994,
[Figure 13, event 21] no such earthquakes were known to have
occurred on the Garlock fault, though there were several
historical
events for which a Garlock fault
source was possible \cite{mcgill}.

The July 1992 event was the larger of
the two recent events, at $M_w=5.3$. This was the largest earthquake
associated with the Garlock fault since the June 10,
1988, $M_L=5.4$ earthquake that occurred several km north of
the Garlock, about 20 km east of its intersection with
the San Andreas fault \cite{mcgill}.
Prior to the 1988 event, the most
recent earthquakes local to the Garlock fault were
two historical events occurring in 1916: a $M5.5$ event 45 km
north of the eastern end of fault, in the Quail mountains
[Toppozada et al., 1978] and a $M5.2$ quake at the western end of the
fault, for which the San Andreas may be responsible.
The July 11, 1992, $M_w 5.3$ Garlock earthquake was clearly related
to and possibly triggered by the sudden changes in the
regional stress field caused by Landers. The 1992 event and
the October 19, 1994, $M_w 4.0$ earthquake lie on
either side of the midpoint
of the Garlock (near the city of Rand), which marks a
change in strike, seismic and aseismic behavior, and
geology \cite{astiz}. The two events lie on either side
of an en-echelon fault step-over near Rand and Koehn lake,
which McGill
and Sieh [1991] argue divides the fault into a western and
an eastern segment.

While the western segment of the Garlock Fault has manifested
continuous low level seismicity and demonstrable creep during
the last several decades, the eastern segment has had only a
few small earthquakes, and no observed creep \cite{astiz}.
The $M_w5.3$ 1992 event, which took place within two weeks of the
Landers mainshock, occurred on the western segment very near
the en echelon step--over, 
at a depth of 11 km [Figure 13, Table V, event 8].
 This event was moderate in
size, with a moment of $M_o=9.44\pm2.29\times10^{23}$ (from our
long--period solution),
but extremely short in source duration, which yields an
unusually high stress-drop of about $1044\pm253$ bars.
Broadband and long--period waveform fits for the July 11, 1992,
Garlock event are shown on Figures 14a and 14b, respectively.
The broadband modeling yields a lower moment estimation,
thus a slightly lower stress-drop
of $840\pm316$ bars. Error associated with moment determination
is greater for the broadband records, which translates into higher
error in the stress-drop estimation.

 The $M_w4.0$ 1994
event occurred on the eastern segment of the Garlock, also near
the en-echelon stepover, and had a stronger thrust component to its motion
[Figure 13, event 21],
and a depth of about 8 km.
The stress--drop is lower than that obtained for the
earlier event, but nonetheless high: $192\pm90$ bars for the
long--period solution. The presence of these
argueably triggered, rare high stress--drop events
on a seismically quiescent fault suggests that small patches of
the fault may rupture energetically, in the first case
at fairly great depth within the crust.
 This further suggests that 
the Garlock may be storing strain, especially near the 
step-over which marks a transition from creeping to
locked behaviour.

\subsection{Summary}

Since duration and moment are routinely computed for each
event we study, we infer stress-drops for these events,
assuming a circular fault. Stress--drops appear to
vary systematically with location, with respect to
previous seismicity or rupture, and  in the case of events in
the Transverse ranges only, with respect to
depth [Figure 15].
Our event sample size is  small in number for any given
region,
yet the events studied here are of moderate size (on average $M\sim4.2$)
thus associated with more energy release than smaller
(and more numerous) events.

We have observed the following for events within the ECSZ:

\begin{itemize}

\item
Joshua Tree events occurred in a historically active
region, and while the sequence was relatively sustained
given the mainshock size, average stress-drops
are relatively low (30 bars) compared to aftershocks
from the Landers sequence both north and south of the
Pinto Mountain fault [Figure 10].

\item
Almost 76\% of total aftershock energy post-Landers was released
south of the mainshock epicenter in the `Southern Landers' area,
yet stress-drops for  these events
are about 50\% lower, on average, than stress-drops for events
north of the Pinto Mountain fault
(i.e.,  67 bars for Southern Landers, and 95 bars for
on-fault and off-fault activity North of the Pinto Mountain
Fault, omitting Garlock events; see Figure 10).

\item
Regions active during the Joshua Tree 
sequence form a stress-drop low during the `Southern Landers'
sequence, and $M>4$ events there were not numerous.
 This suggests that while the Landers mainshock may
have 'recharged'  aftershock activity in the Joshua Tree
region [Hauksson, 1994],  moment-release and stress-drop in the
region remained low.

\item
Heterogeneous and high stress-drop aftershocks occurred along the newly
recognized Landers-Kickapoo fault,  associated with smaller
surficial slip on the Landers fault relative to Johnson Valley
fault (JVF)  and
Homestead valley faults (HVF)  
and lack of through-going dextral rupture across the JVF/HVF
stepover.  High stress-drop events in this area may be
related to the presence of the immature Landers fault and an
incomplete connection between the Johnson Valley and Homstead
Valley fault systems.

\item
High stress-drops in the Pisgah-Calico region might suggest
high applied shear-stresses on North-South planes, while
relative stress-drops in historically active Barstow were
appreciably lower, much like aftershocks in the Joshua Tree region.

\item
In the immediate aftermath of the Landers event, a large,
rare, high-stress-drop event ocurred on the historically
quiescent Garlock fault. Two years later
a second event occurred near the stepover from the creeping western 
segment to the 'locked' eastern strand of the
fault. The presence of these two events on a historically
aseismic fault suggests that small patches of a quiescent
fault may rupture very energetically, and also that the
Garlock may be storing strain, especially at the stepover
which marks a transition from creeping to locked behavior.

\item
In contrast to aftershocks from the Big Bear sequence,
Landers aftershocks are in general shallower [Jones and Helmberger, 1996].
While Landers and Big Bear events are all moderately high stress--drop
(on average, 70 bars for the Landers events, 100 bars for Big Bear,
see Figure 15), events occurring in the eastern Transverse ranges
are generally higher stress-drop, and show a strong correlation between
high stress-drop and greater event depth.  Like events in the Transverse
ranges, however, high 
stress drops for Landers events appear to correlate with activity on 
immature or low-slip faults.

\end{itemize}

\section{Conclusions}
The Landers mainshock
and related events altered the tectonic landscape
and stress budget of Southern California in ways not yet
fully assessed. The Landers earthquake itself involved surface
rupture and displacement on six separate faults, including rupture
south of the Pinto Mountain fault on the Eureka Peak fault.
Aftershocks and triggered events occurred as far away as Mammoth
Lakes, California,  and Little Skull Mountain, Nevada \cite{hill}, 
and included the complex $M6.5$
Big Bear mainshock, and several unusual earthquakes on the Garlock
fault.
 
For the Landers sequence, stress--drops of
events located at some distance from the Landers rupture
are higher than those located on the faults involved in
the mainshock,  with the exception of aftershocks on
the juvenile Kickapoo (Landers) fault. Rupture on this
fault segment was complicated, and displacement
may have been accommodated across a number of
subsidiary or discontinuous fault traces. The fact
that the Kickapoo fault had some of the lowest measured
surface displacements
during the Landers mainshock lends credence to
this idea.

Aftershock stress drop patterns often show a 
low associated with the mainshock fault--plane. We
observe an analoguous phenomenon in the low stress-drops
recorded for previously active regions of the strike-slip
system comprising Southern and Northern Landers.
Work by Smith and Priestly [1993] on the 1984 Round Valley, California,
earthquake showed an aftershock stress-drop minimum on the fault-plane,
suggesting nearly complete stress-release  in the ruptured area.
Consistent with their work, and with
theories of fault rupture and asperity [Madariaga, 1973], 
is our observation that stress drops are 
relatively higher 
off-fault
and around the edges of the rupture trace.

High stress-drops have been associated with
long earthquake recurrence times [Kanamori and Allen, 1986;
Scholz et al., 1986], which may in turn be related to
low slip rates on locked, discontinuous, or youthful
faults.  In the aftermath of the Landers quake, an
unusual, deep, high stress-drop event was triggered on the Garlock
fault, which has not experienced any large earthquakes 
within the period of historical record, though  scarps
and offset features suggest it has produced large quakes
in the past. Here again is an example of a quiescent
fault producing high stress--drop events. 

Aftershocks South of the Pinto Mountain fault
occurred in a region associated with high rates of
post--seismic deformation, like those in the
Barstow region \cite{shen}. Lower stress--drop
aftershocks seem to occur in regions which previously
experienced the most local moment release; i.e.,
near the Eureka Peak fault, and near the Joshua Tree
mainshock epicenter.

\pagebreak 
\begin{table}
\begin{center}
\centerline{Table III: Joshua Tree Aftershocks, $M_w>4$}
\begin{tabular}{cccccccccc}
  \hline 
   &&&&&&&\multicolumn{3}{c}{Location} \\ \cline{8-10}
    No.& Date    &  $M_w$ & $\tau$, $\triangle\sigma$ & $\theta$ & $\delta$ & $\lambda$ & Depth & Latitude & Longitude \\
       &         &        & s, bars &&&&  km &  $^o$N  &  $^o$W \\
\hline   
  1.& 92042302  &  4.3  & 0.45,74 & 170 & 82 & 154 &  12 &  33.94 & 116.33 \\
  2.& 92042318  &  4.0  & 0.45,19 &  334 & 50 & 130 &  8  &  33.97 & 116.29 \\
  3.& 92042606  & 4.5  & 1.15,10 &  354 & 60 & 224 & 8   &  33.92 & 116.33 \\
  4.& 92042703  & 4.3  & 1.10,4 & 156 & 74 & 162 & 5 & 33.91 & 116.34 \\
  5.& 92050416  & 4.8 & 0.80,70 & 170 & 80 & 190 & 14 & 33.92 & 116.32 \\
  6.& 92050602  & 4.5 & 0.90,21 & 356 & 72 & 238 & 11 & 33.92 & 116.32 \\
  7.& 92051202  & 4.3 & 0.80,13 & 352 &  70 & 184 & 8 &  33.96 & 116.28 \\
  8.& 92051815  & 4.7 & 0.80,55 & 346 &  66 & 224 & 11 &  33.95 & 116.35 \\
  9.& 92061100  & 4.4 & 1.10,6 &  172 & 74 & 196 &  9  &  34.21 & 116.30 \\
\hline   
\end{tabular}
\label{lantab2}
\end{center}  
\end{table}

\pagebreak

\begin{table}
\begin{center}
\small
\centerline{Table IV: Landers Events, South of Pinto Mountain Fault}
   \begin{tabular}{cccccccccc}
   \hline
   &&&&&&&\multicolumn{3}{c}{Location} \\ \cline{8-10}
    No. & Date  &  $M_w$ & $\tau$, $\triangle\sigma$ & $\theta$ & $\delta$ & $\lambda$ & Depth & Latitude & Longitude \\
        &       &        & s, bars & & & &  km &  $^o$N  &  $^o$W \\
   \hline
  1.& 92063011  &  4.2  & 0.35,85 &  353 & 51 & 215 &  14 &  34.07 & 116.45 \\
  2.& 92063014  &  5.1  & 1.0,90 &  350 & 45 & 200 &  7  &  34.00 & 116.37 \\
  3.& 92070612  & 4.2  & 0.60,28 &  330 & 76 & 182 & 8  &  34.09 & 116.33 \\
  4.& 92070619  & 4.3  & 0.60,28 &  160 & 62 & 208 & 9  &  34.07 & 116.34 \\
  5.& 92071002  & 3.9  & 0.50,10 & 132 & 70 & 218 & 11 & 34.12 & 116.40 \\
  6.& 92072418  & 4.9 & 1.0,52  & 351 & 80 & 173 & 8 & 33.90 & 116.28 \\
  7.& 92072504  & 4.7 & 1.0,25  & 2 & 76 & 238 & 8 & 33.94 & 116.30 \\
  8.& 92072818  & 4.7 & 1.0,25  &  310 &  40 & 100 & 5 &  34.09 & 116.37 \\
  9.& 92081106  & 4.1 & 0.40,45 &  336 & 80 & 170 &  8 &  34.06 & 116.37 \\
  10.& 92081508 & 4.5 & 0.35,346 & 338 & 58 & 190 &  6 & 34.088 & 116.403 \\
  11.& 92090912  &  4.2 & 0.50,38 & 112 & 62 & 110 & 8 &  33.94 & 116.33 \\
  12.& 92091508  &  5.2 & 1.50,30 & 156 & 76 & 188 & 8 &  34.09 & 116.35 \\
  13.& 93082101 &  4.5 & 0.65,60 & 208 & 54 & 278 & 9 & 34.010 & 116.32 \\
  14.& 94080715 &  3.7 & 0.60,4 & 352 & 64 & 184 & 8 & 33.99 & 116.28 \\
  15.& 94081508 &  3.8 & 0.25,76  & 146 & 64 & 240 & 9 & 33.81 & 116.20 \\
   \hline
\end{tabular}
\label{lantab3}
\end{center}
\end{table}

\pagebreak

\begin{table}
\begin{center}
\small
\centerline{Table V: Landers Events, North of Pinto Mountain Fault}
   \begin{tabular}{cccccccccc}
   \hline
   &&&&&&&\multicolumn{3}{c}{Location} \\ \cline{8-10}
   No. & Date &  $M_w$ & $\tau$, $\triangle\sigma$ & $\theta$ & $\delta$ & $\lambda$ & Depth & Latitude & Longitude \\
       &      &        &  s, bars &&&&  km &  $^o$N  &  $^o$W \\
   \hline
  1.& 92063012  & 4.0  & 0.30,84 & 342 & 50 & 254 &  9 &  34.32 & 116.45 \\
  2. & 92063017 &  4.1  & 0.40,46 &  156 & 74 & 222 &  8  &  34.64 & 116.66 \\
  3.& 92070107  & 5.2  & 0.50,515 &  194 & 76 & 160 &  7 &  34.33 & 116.46 \\
  4.& 92070510  & 4.5 & 0.80,25 & 331  & 80 & 169 & 8 &  35.03 & 116.97 \\
  5.& 92070521  &  5.4  & 1.50,71 &  344 & 70 & 142 &  8  &  34.58 & 116.32 \\
  6.& 92070522  &  4.4  & 0.70,25 &  336 & 64 & 140 &  8  &  34.57 & 116.33 \\
  7.& 92070802  & 4.6  & 0.50,140 & 162 & 66 & 156 & 8 & 34.57 & 116.30 \\
  8.& 92071118 & 5.3 & 0.55,1044 & 296 & 58 & 164 & 11 & 35.21 & 118.07 \\
  9.& 92071500 & 3.9  & 0.40,30 & 20 & 68 & 186 & 6 & 34.33 & 116.46 \\
  10.& 920720040  & 3.9 & 0.60,9 &  320 & 84 & 224 & 8 &  34.20 & 116.45 \\
  11.& 920720044 & 4.4  & 0.80,16 &  358 & 82 & 204 & 7 &  34.96 & 116.95 \\
  12.& 92072013 & 4.5  & 0.60,63 & 348 & 71 & 183 & 5 &  34.98 & 116.96 \\
  13.& 92072407 & 3.8 & 0.30,38 &  344 & 60 & 260 & 11 &  34.48 & 116.50 \\
  14.& 92080522 & 4.6 & 0.60,80 & 146 & 82 & 210 & 6 & 34.98 & 116.97 \\
  15.& 92080815 & 4.3 & 0.40,86 & 168 & 64 & 146 & 8 & 34.37 & 116.45 \\
  16.& 92083109 & 4.2 & 0.35,78 & 154 & 90 & 160 & 12 &  34.50 & 116.43 \\
  17.& 92100207 &  4.6 & 0.35,250 & 189 & 83 & 313 & 5 & 34.61 & 116.64 \\
  18.& 92101112 &  4.4 & 0.60,52 & 170 & 64 & 140 & 8 & 34.93 & 116.82 \\
  19.& 94061616 &  4.7 & 1.2,13 & 148 & 61 & 193 & 5 & 34.267 & 116.40 \\
  20.& 94080121 &  4.4 & 0.40,126 & 360 & 78 & 202 & 14 & 34.633 & 116.523 \\
  21.& 94101900 & 4.2 & 0.22,190 & 126 & 50 & 150 & 8 & 35.51 & 117.48 \\
   \hline
   \end{tabular}
\end{center}
\end{table}

\pagebreak

\begin{acknowledgements}
Contribution number 5800, Division of Geological and Planetary
Sciences, California Institute of Technology, Pasadena, CA 91125.

\end{acknowledgements}


\clearpage


\begin{figure}
\centerline{\vbox{
\psfig{figure=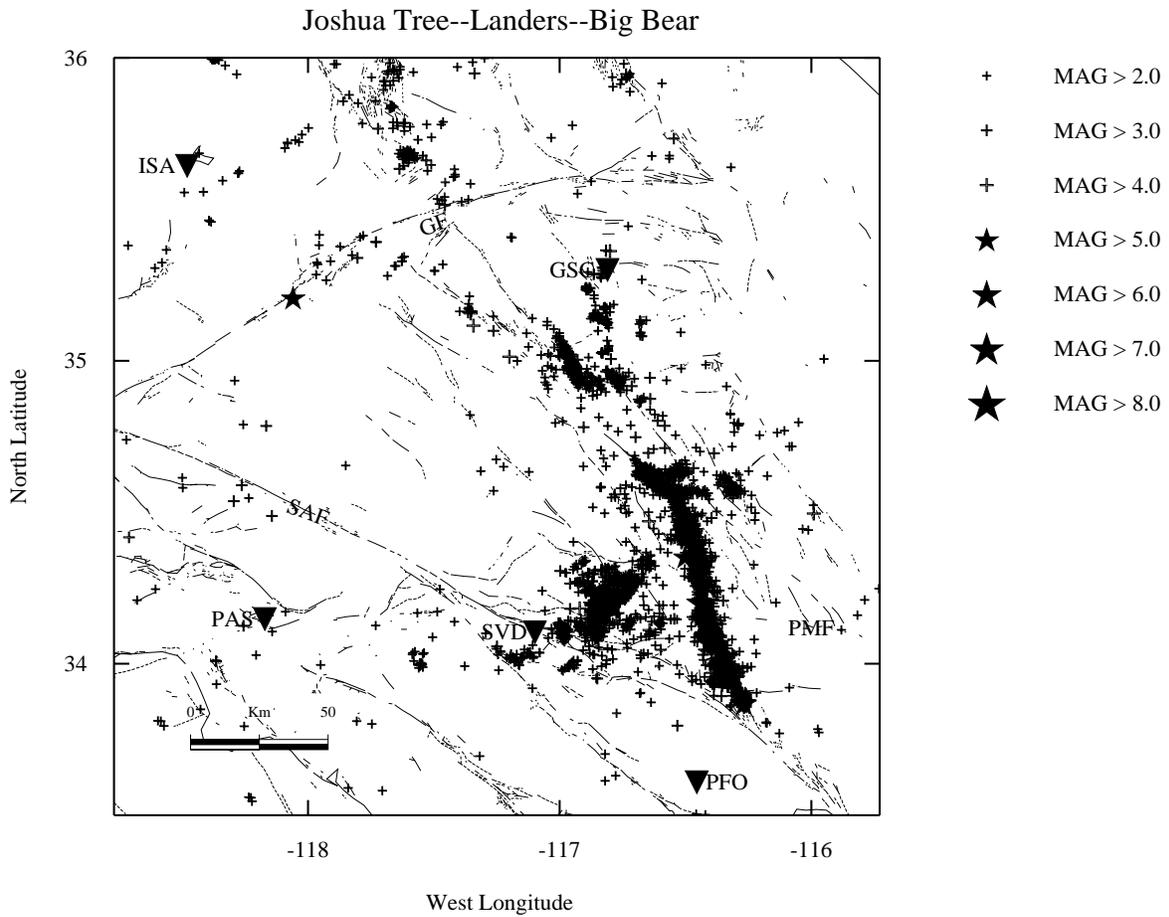,width=6.00in}}}
\caption{Location map showing main events and aftershocks from the
Joshua Tree, Landers and Big Bear sequences. Map covers
seismicity from April 23, 1992, to December 31, 1992. Faults
are indicated as follows: SAF (San Andreas fault), GF (Garlock
fault), PMF (Pinto Mountain fault).}
\end{figure}

\clearpage


\begin{figure}
\centerline{\vbox{
\psfig{figure=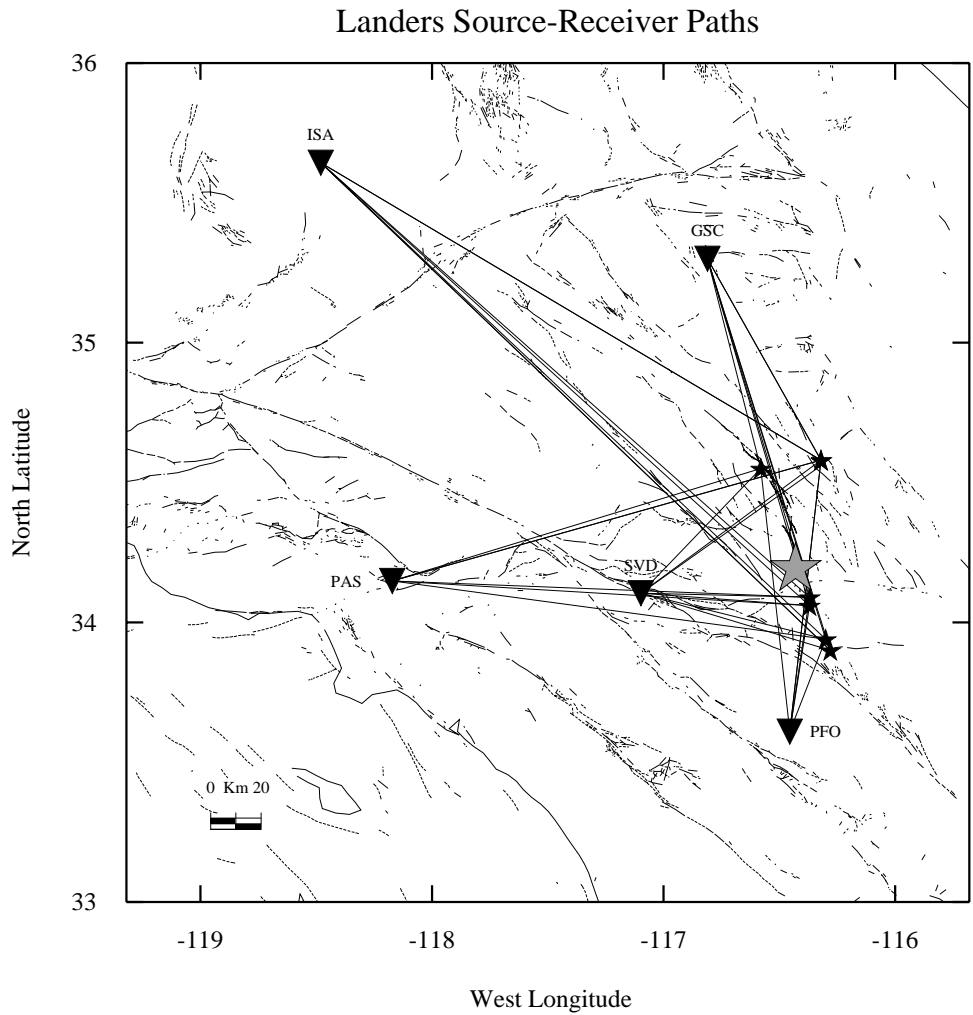,width=5.00in}}}
\caption{Source--receiver paths for the profiles used in
source modeling, and in the construction and testing of
the Mojave Model. Stations GSC, PFO and SVD were used
primarily in the estimation of source mechanisms for Landers
and Joshua Tree events. Stations ISA and PAS were included
as needed, to create a robust solution in cases where the
solution appeared unstable. Source--event paths for
stations GSC and PFO were used in the development of the
Mojave model (Table I).}
\end{figure}

\clearpage


\begin{figure}
\centerline{\vbox{
\psfig{figure=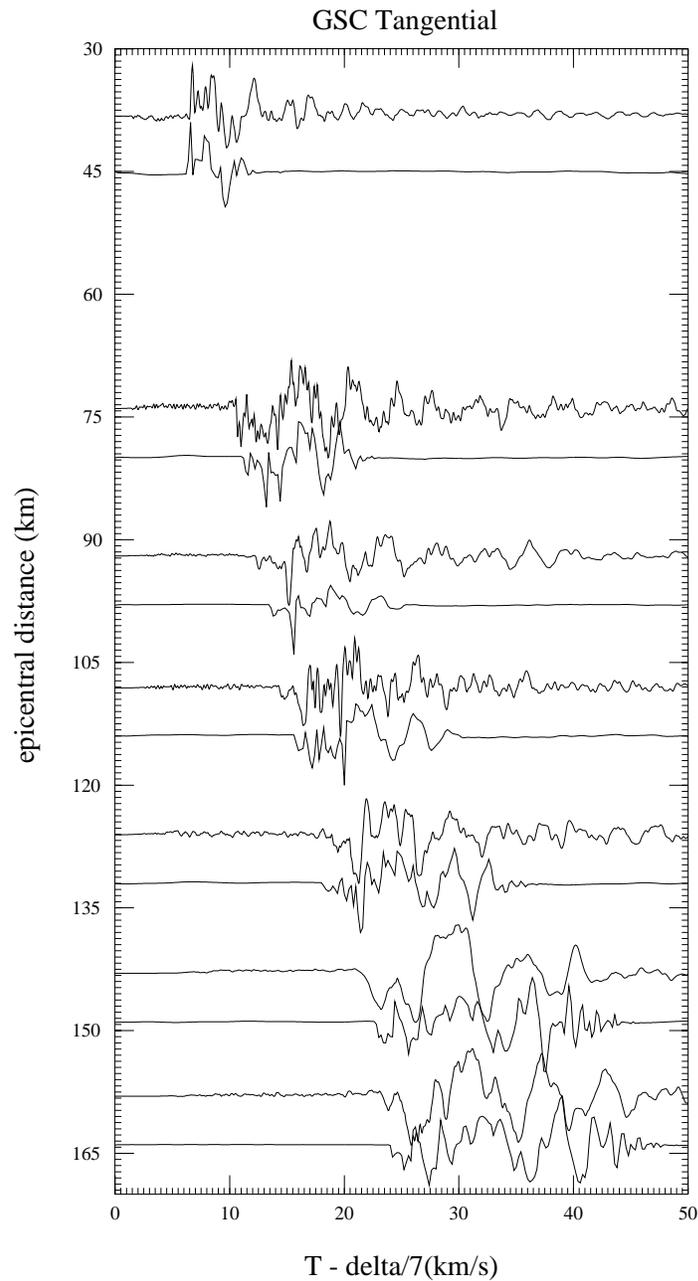}}}
\caption{
Profile of Landers data and modeling for the tangential component of
displacement recorded at
station GSC. This profile ranges north to south, with
source--receiver distances ranging from 40 to 160 km and
source depths between 8 and 11 km, roughly average for this sequence.
Source mechanisms used in the modeling 
are computed using the
methods discussed in text.
Records are modeled and shown broadband;
observed displacement records are shown in bold
line above synthetics. Synthetics are generated using the
Mojave model (this paper, Table I) and the frequency--wavenumber method.
}
\end{figure}

\clearpage


\begin{figure}
\centerline{\vbox{
\psfig{figure=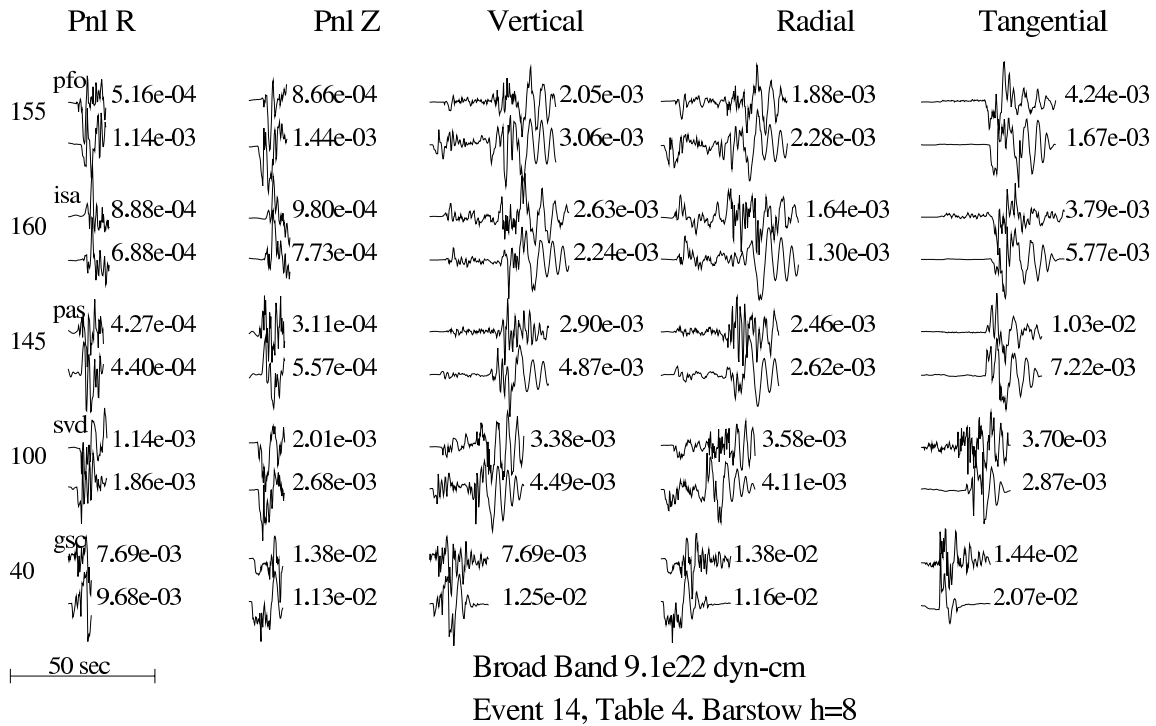,angle=270,width=6.00in}}}
\caption{Broadband modeling for the August
5, 1992 22:22 (Barstow) aftershock. Source depth was estimated
at between 5 and 8 km by cycling through synthetics appropriate to
source depths from 2 to 17 km, and finding a minimum error
solution. Event duration was estimated first by measuring the
direct pulse, then by the energy method described in this paper.
Synthetics are generated using the F-K method and
the Mojave model. This plot shows waveform fits assuming a depth
of 8 km; the next plot shows the depth of 8 km.}
\end{figure}

\clearpage


\begin{figure}
\centerline{\vbox{
\psfig{figure=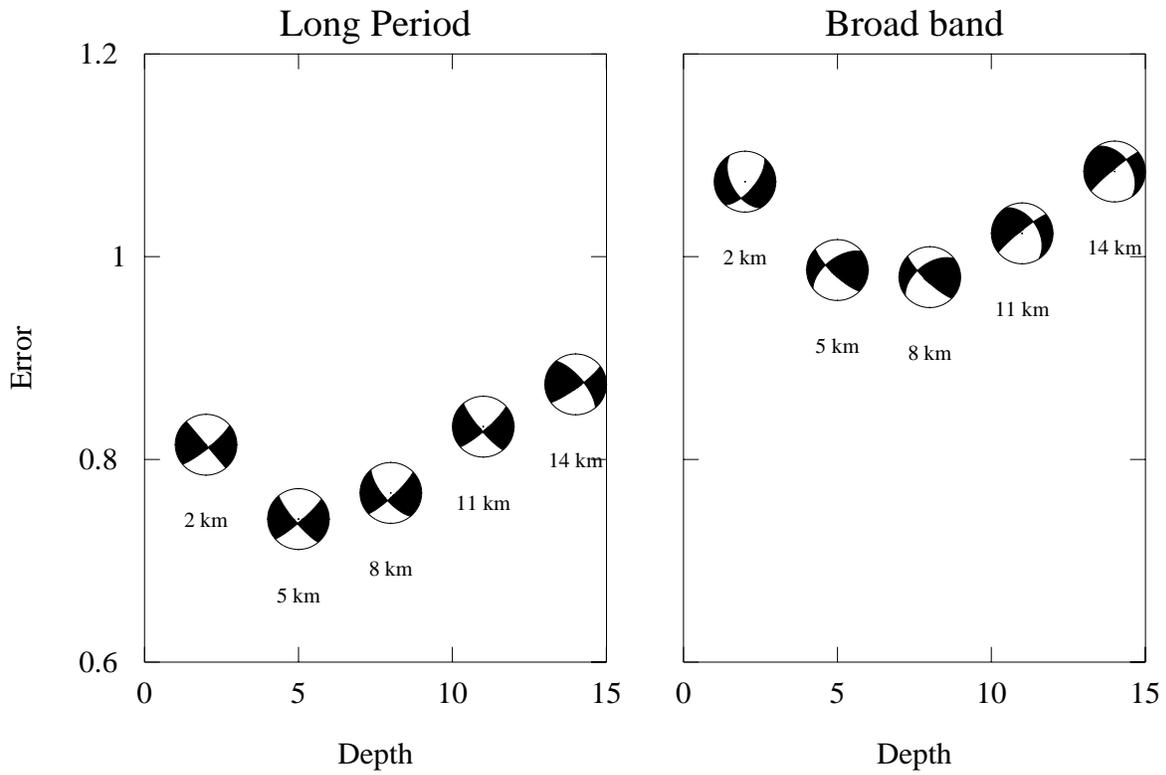,angle=270,width=6.00in}}}
\caption{
Error space for the August 5, 1992, 22:22 (Barstow)
event. Source depths are indicated across the bottom
of the plot, and error on the vertical
axis. The left--hand panel shows error from the Long--period
solution, and the right--hand panel shows error from the
broadband solution. Focal spheres appropriate to each depth
indicate data points; note that long--period focal spheres show
more consistency.
}
\end{figure}

\clearpage


\begin{figure}
\centerline{\vbox{
\psfig{figure=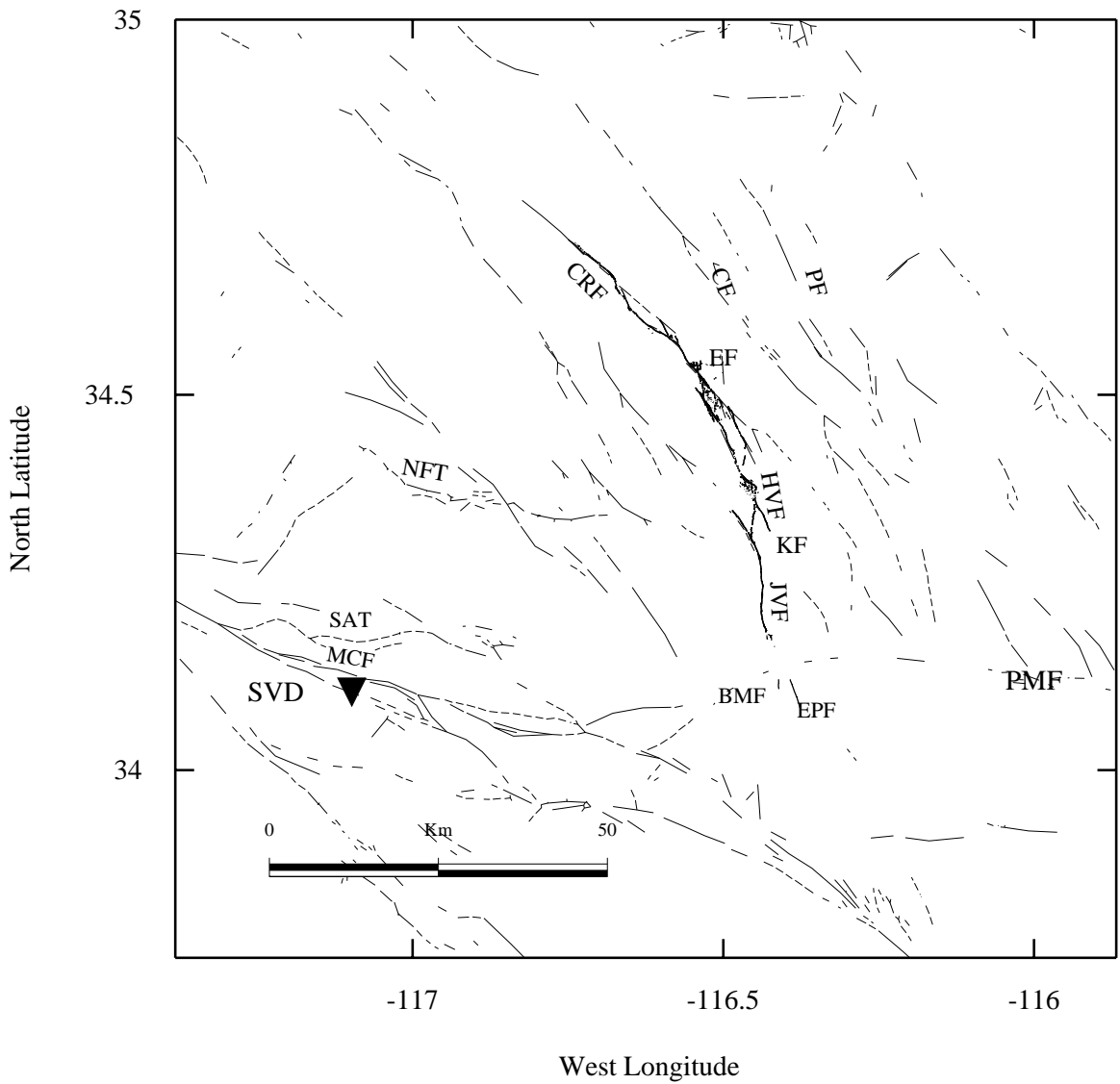,width=6.00in}}}
\caption{
Location map showing faults active during the
Joshua Tree, Landers and Big Bear sequences.
Faults are indicated as follows, clockwise from lower left: MCF, Mill Creek
fault; SAT, Santa Ana Thrust;  NFT, North Frontal Thrust; CRF,
Camp Rock fault; CF, Calico Fault; PF, Pisgah fault; EF, Emerson
fault; HVF, Homestead valley fault; KF, Kickapoo (Landers) fault;
JVF, Johnson Valley fault; PMF, Pinto Mountain fault; EPF, Eureka
Peak fault and
BMF, Burnt Mountain fault. The Garlock fault is shown
on Figure 1.
}
\end{figure}

\clearpage


\begin{figure}
\centering
\centerline{\vbox{
\psfig{figure=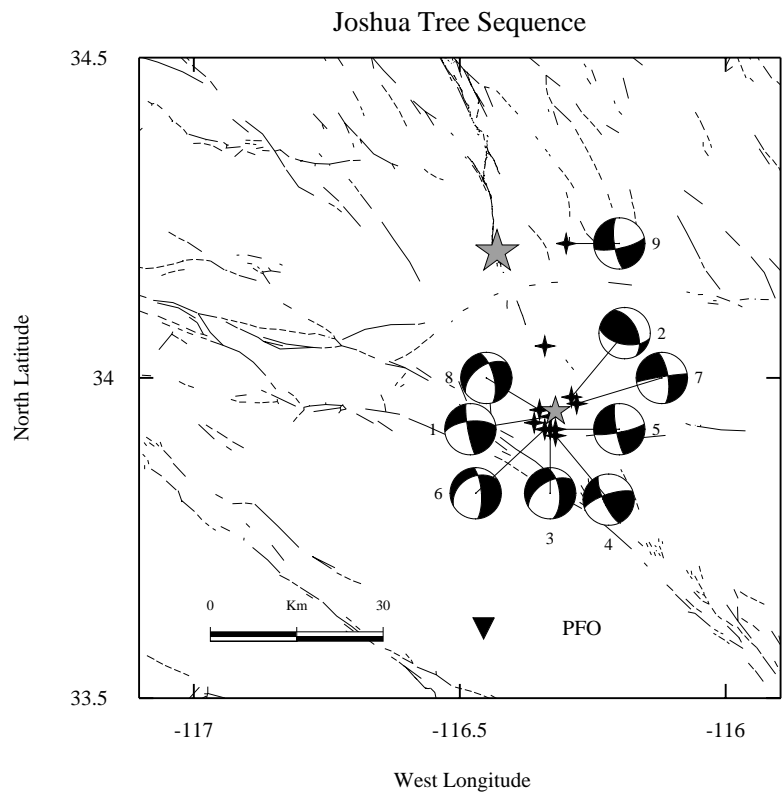,width=4.00in}}}
\caption{
(a) Location map showing Joshua Tree aftershocks.
Aftershocks are numbered in order of occurrence, and are listed
in this order in Table III Large filled star is location of
Landers mainshock; small filled star is location of Joshua
tree preshock.
}
\end{figure}

\clearpage
\addtocounter{figure}{-1}

\begin{figure}
\centerline{\vbox{
\psfig{figure=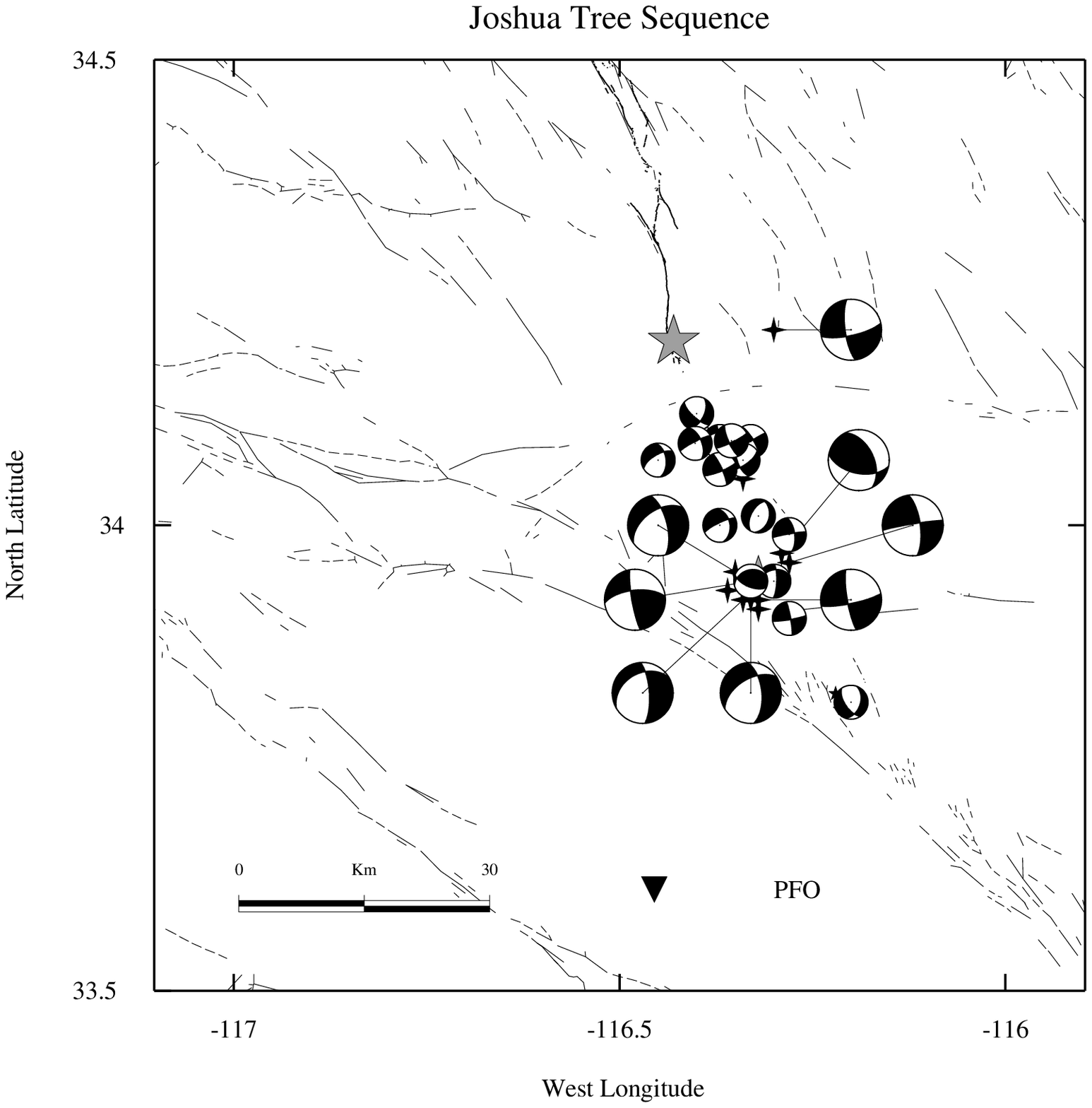,width=4.00in}}}
\caption{
(b) Relative locations of Landers
and Joshua Tree aftershocks. Joshua Tree aftershocks are
indicated with larger spheres; epicentral locations are
crosses. Landers aftershocks in this area are smaller focal
spheres.
Size of focal sphere is not related to event magnitude.
}
\end{figure}

\clearpage


\begin{figure}
\centerline{\vbox{
\psfig{figure=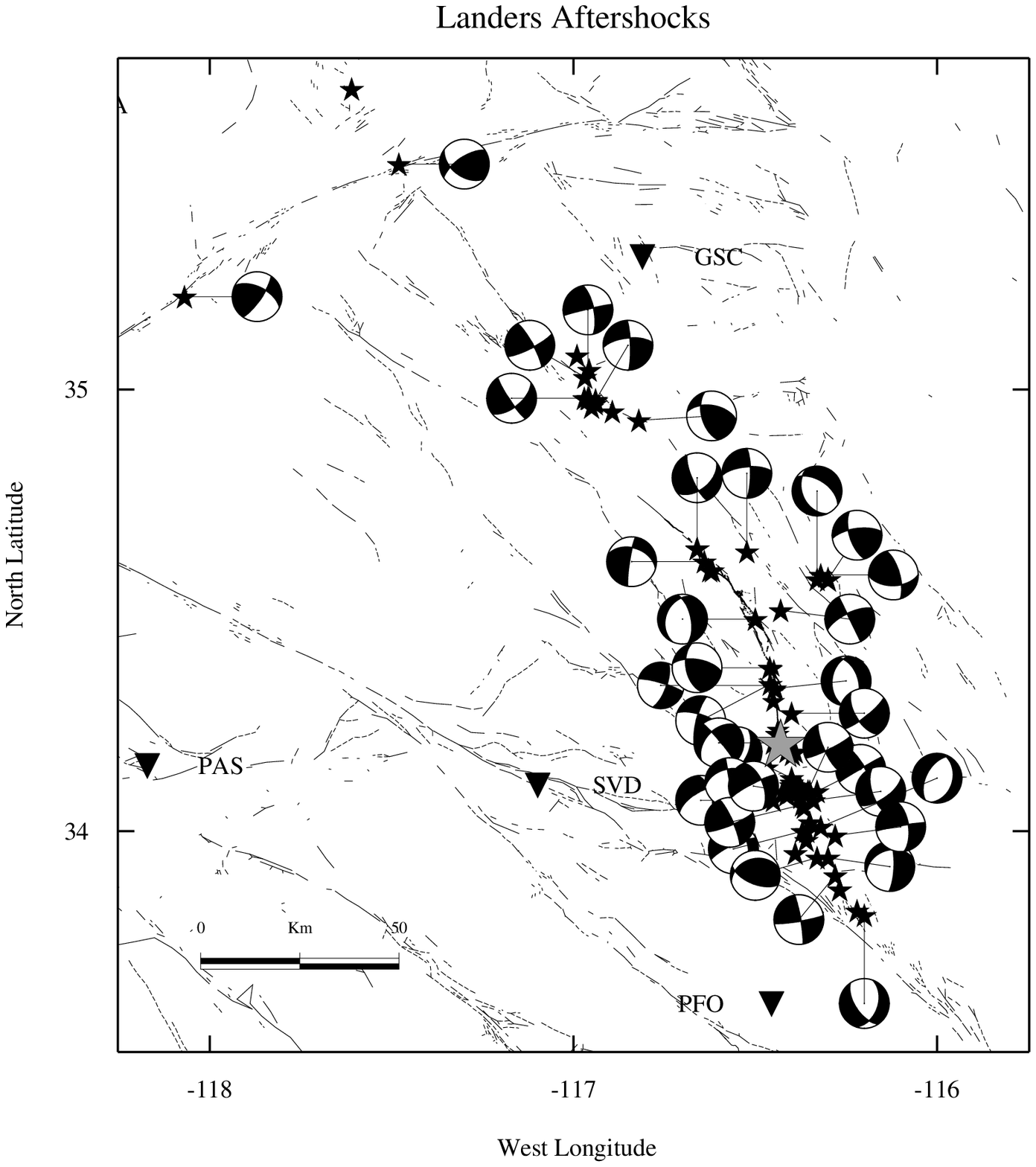,width=6.00in}}}
\caption{
Map of locations and focal spheres for the 34
Landers--related Mojave events discussed here,
including two earthquakes on the Garlock fault.
Epicentral locations are shown as filled (black) stars.
The Landers mainshock is shown as a filled (grey) star. The sequence
shown here includes events occurring from June of 1992 through
October of 1994. These events will be further broken down
and discussed by location and order of occurrence
[i.e.,Figures 11, 13].
}
\end{figure}

\clearpage


\begin{figure}
\centerline{\vbox{
\psfig{figure=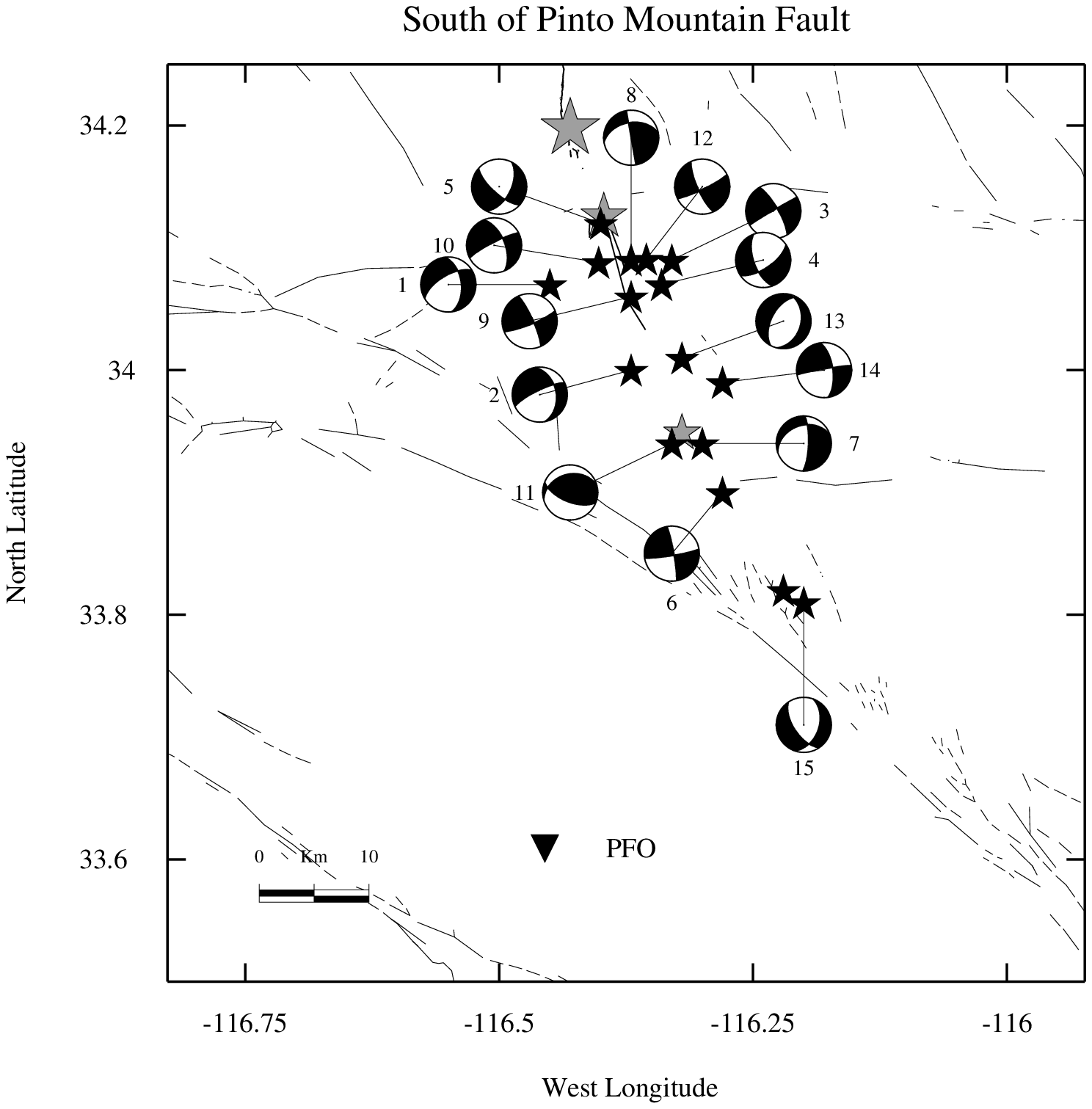,width=4.00in}}}
\caption{
Map showing Landers aftershocks south of
the Pinto Mountain fault. Locations of these aftershocks were
previously shown relative to earlier Joshua Tree aftershocks
(Figure 7b). In this map, the aftershocks are
numbered chronologically, and listed in the same order in
Table IV. The Joshua Tree mainshock is shown
as a small filled (grey) star; Landers mainshock and Southern Landers
subevent are also shown as filled (grey) stars.
}
\end{figure}

\clearpage


\begin{figure}
\centerline{\vbox{
\psfig{figure=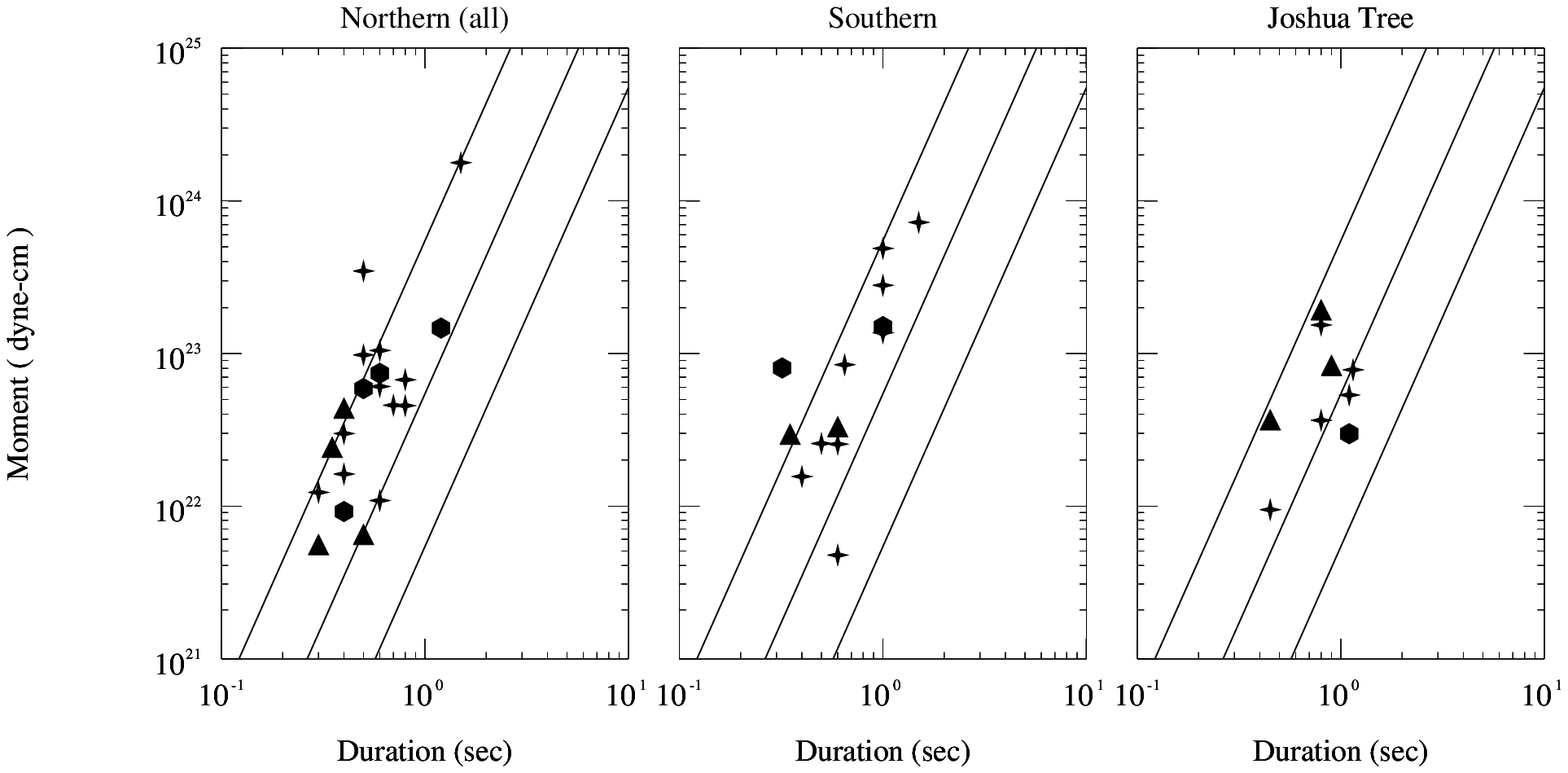,angle=270,width=6.00in}}}
\caption{
Moments versus durations for Joshua Tree aftershocks and
Landers events both north and south of the Pinto Mountain fault.
Event depths are indicated by different symbols:  filled
triangles indicate comparatively ``deep'' events (12 to 17 km);
filled crosses indicate ``intermediate'' depth events (8 to 11 km);
and filled hexagons indicate ``shallow'' events (2 to 7 km).
Lines of constant stress drop are plotted diagonally across
the figure; from top to bottom: 100, 10, and 1 bar(s).
The first panel shows Landers events north the of Pinto
Mountain fault, the second shows events south of the Pinto
Mountain fault and the third shows Joshua Tree events.
}
\end{figure}

\clearpage

 
\begin{figure}
\centerline{\vbox{
\psfig{figure=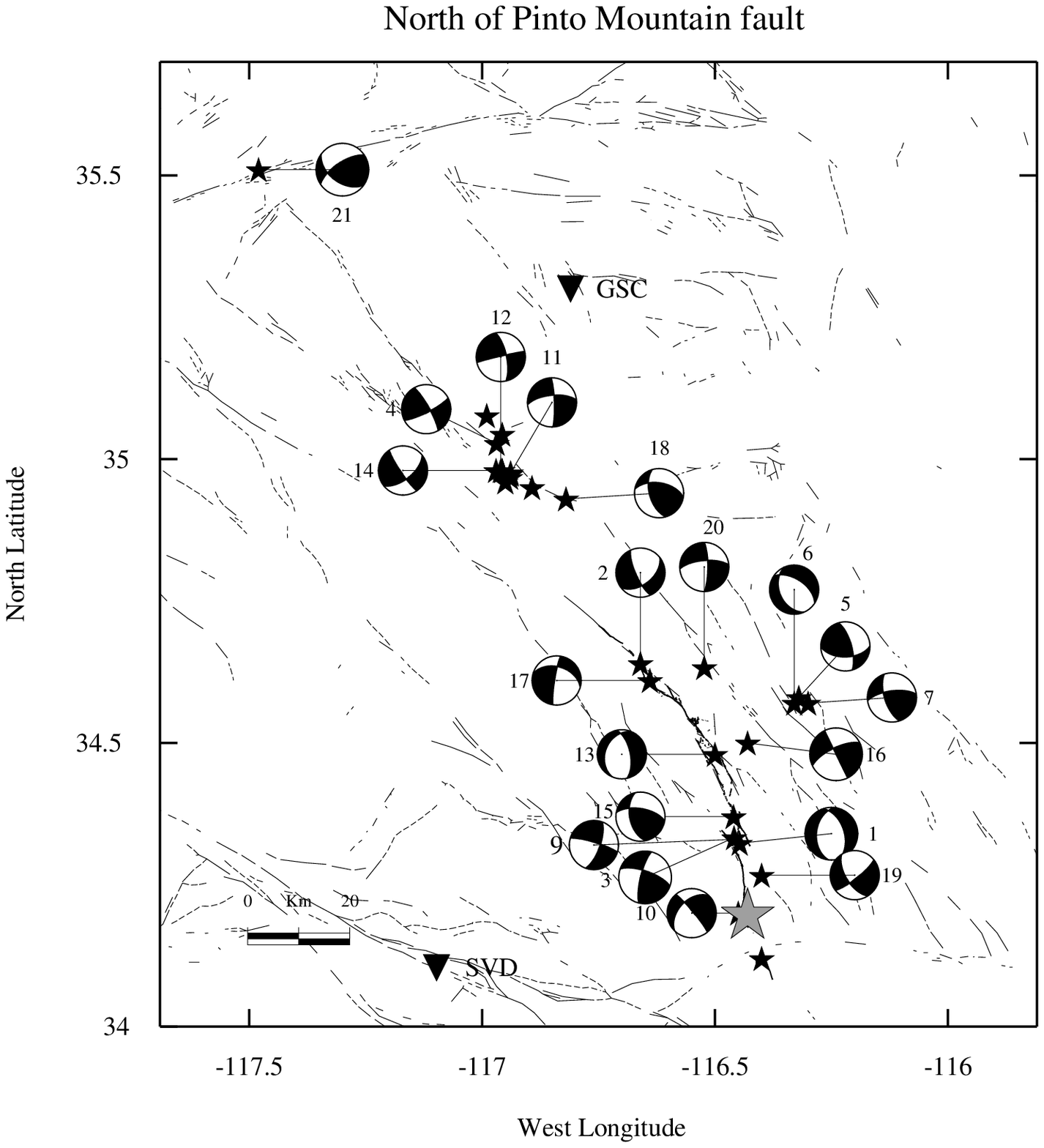,width=4.00in}}}
\caption{
Map showing Landers aftershocks north of the Pinto
Mountain fault, including off-fault clusters at Barstow,
and on the Pisgah and Calico faults. Events are numbered in
the order of occurrence, and listed in this order in
Table V.
}
\end{figure}
 
\clearpage


\begin{figure}
\centerline{\vbox{
\psfig{figure=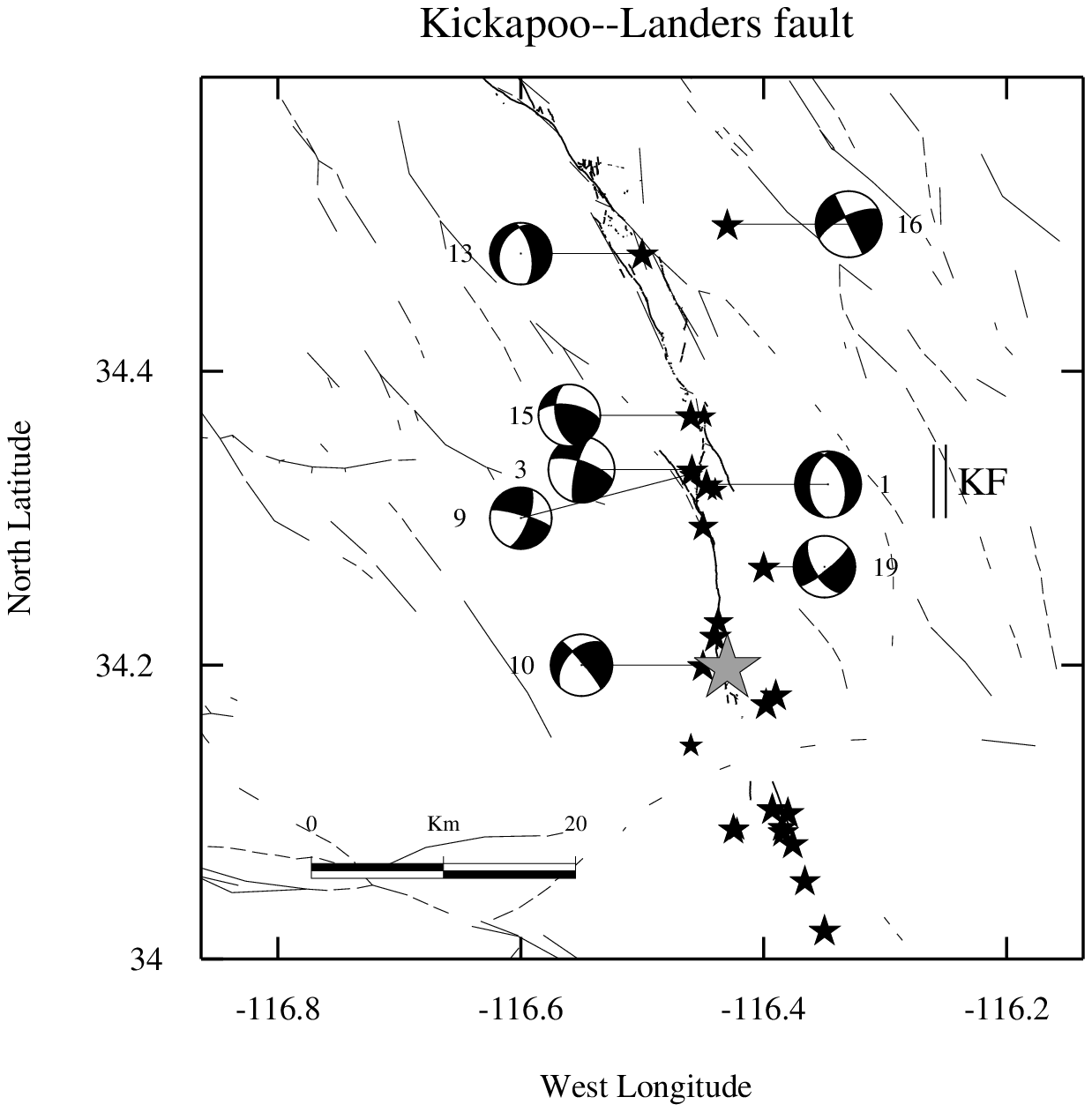,width=4.00in}}}
\caption{
Detail of map from Figure 11, showing
seismicity around the mainshock area (large grey star)
 and Kickapoo fault
(indicated by double line and the letters KF). All
events of $M>4.0$ are shown. Most seismicity south of
the Pinto Mountain fault and around mainshock epicenter
occurred within the first 24 hours.
}
\end{figure}

\clearpage


\begin{figure}
\centerline{\vbox{
\psfig{figure=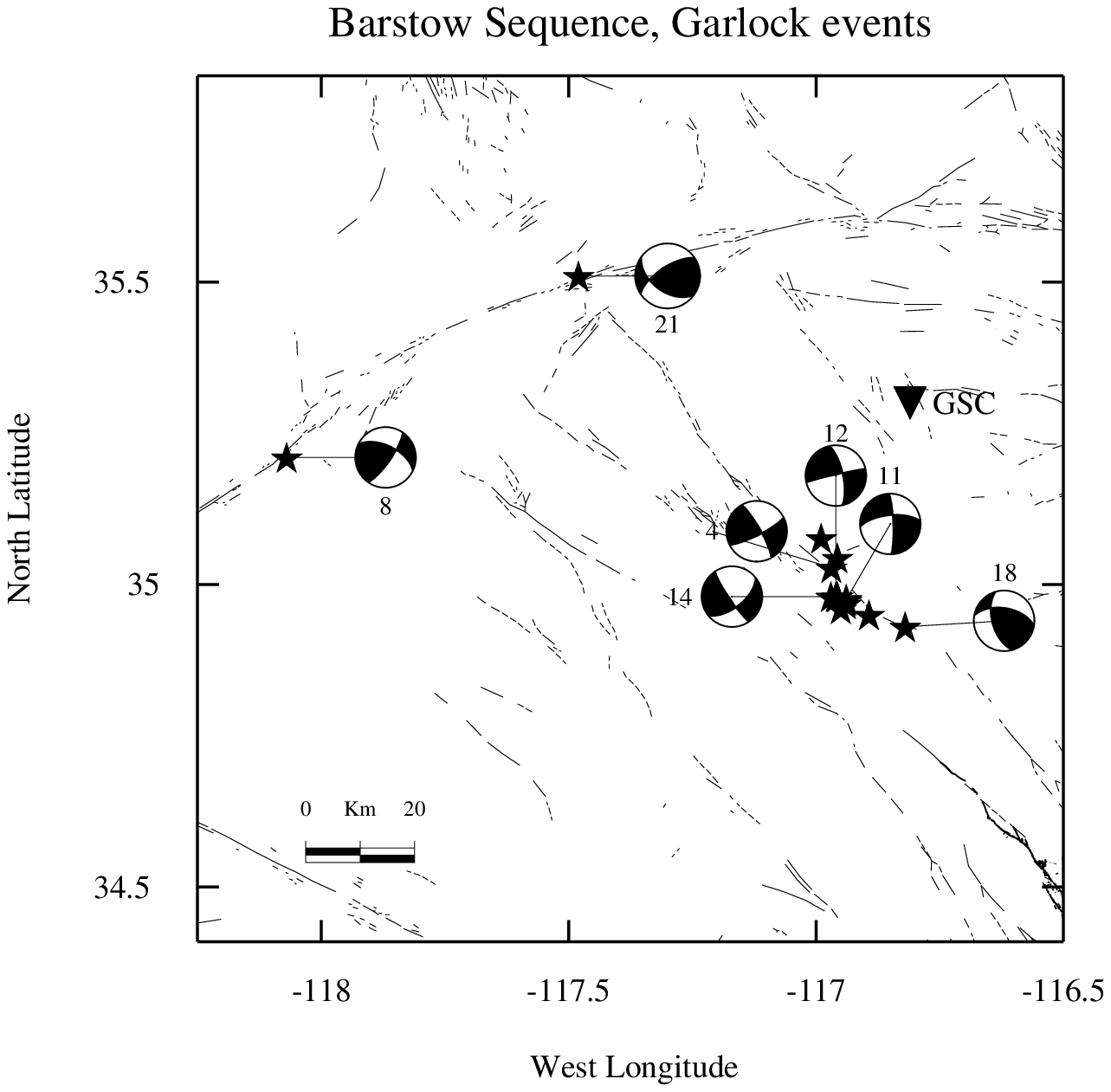,width=4.00in}}}
\caption{
Detail of map from Figure 11, showing
off--fault seismic activity in the Barstow area, and further
north along the Garlock fault. Events are numbered as in
Figure 11 and in Table V.
}
\end{figure}

\clearpage


\begin{figure}
\centerline{\vbox{
\psfig{figure=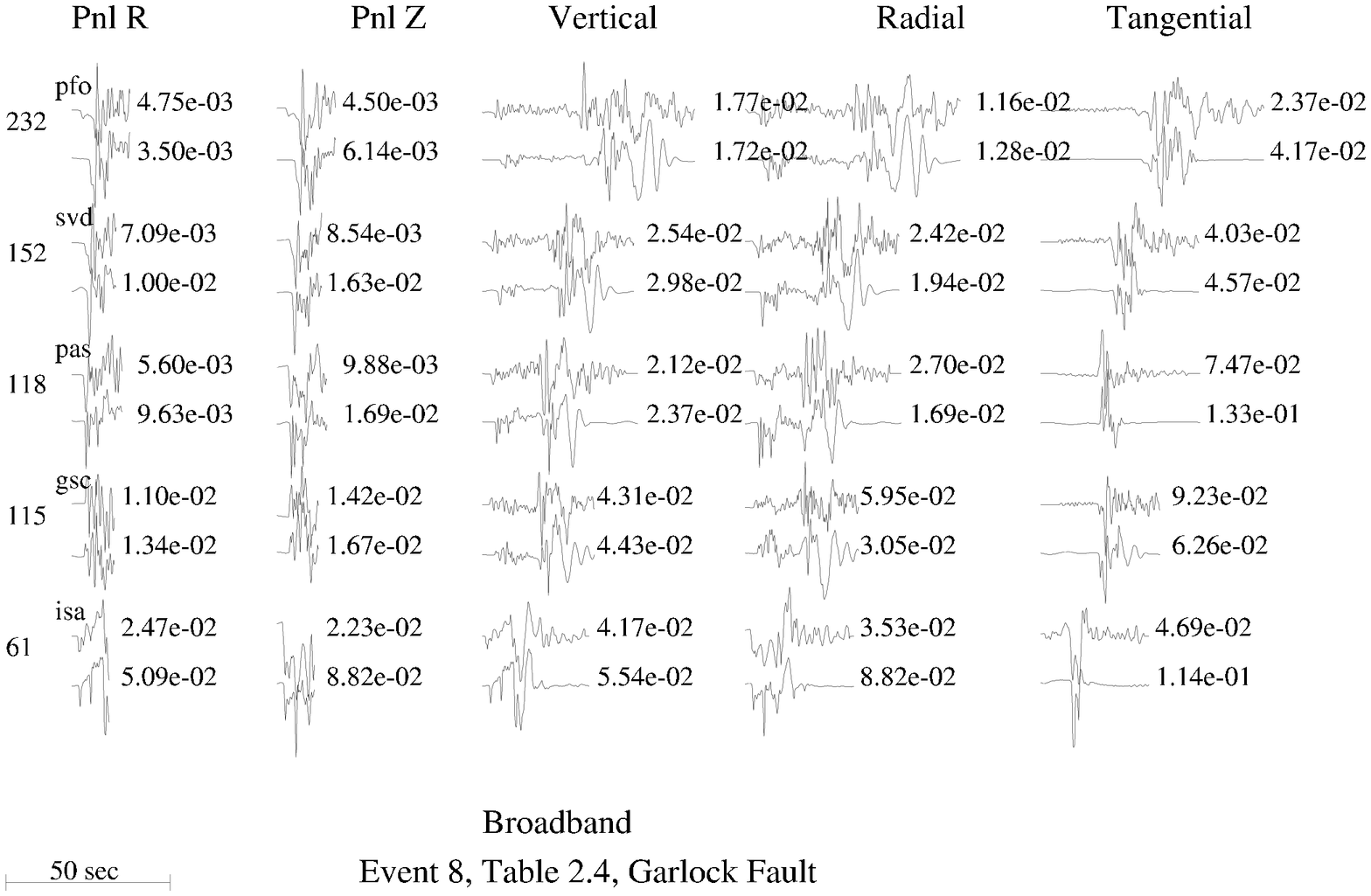,angle=270,width=6.00in}}}
\caption{
(a) Broadband waveform modeling for the $M5.3$ July 11, 1992,
Garlock earthquake. Both the standard Southern California
model (stations PAS, PFO) and Mojave model (GSC, ISA, SVD)
were used in this source estimation. The moment for this
solution is $M_b=7.64\pm2.85\times10^{23}$; the time
function is (0.25, 0, 0.25) s.
}
\end{figure}

\clearpage

\addtocounter{figure}{-1}

\begin{figure}
\centerline{\vbox{
\psfig{figure=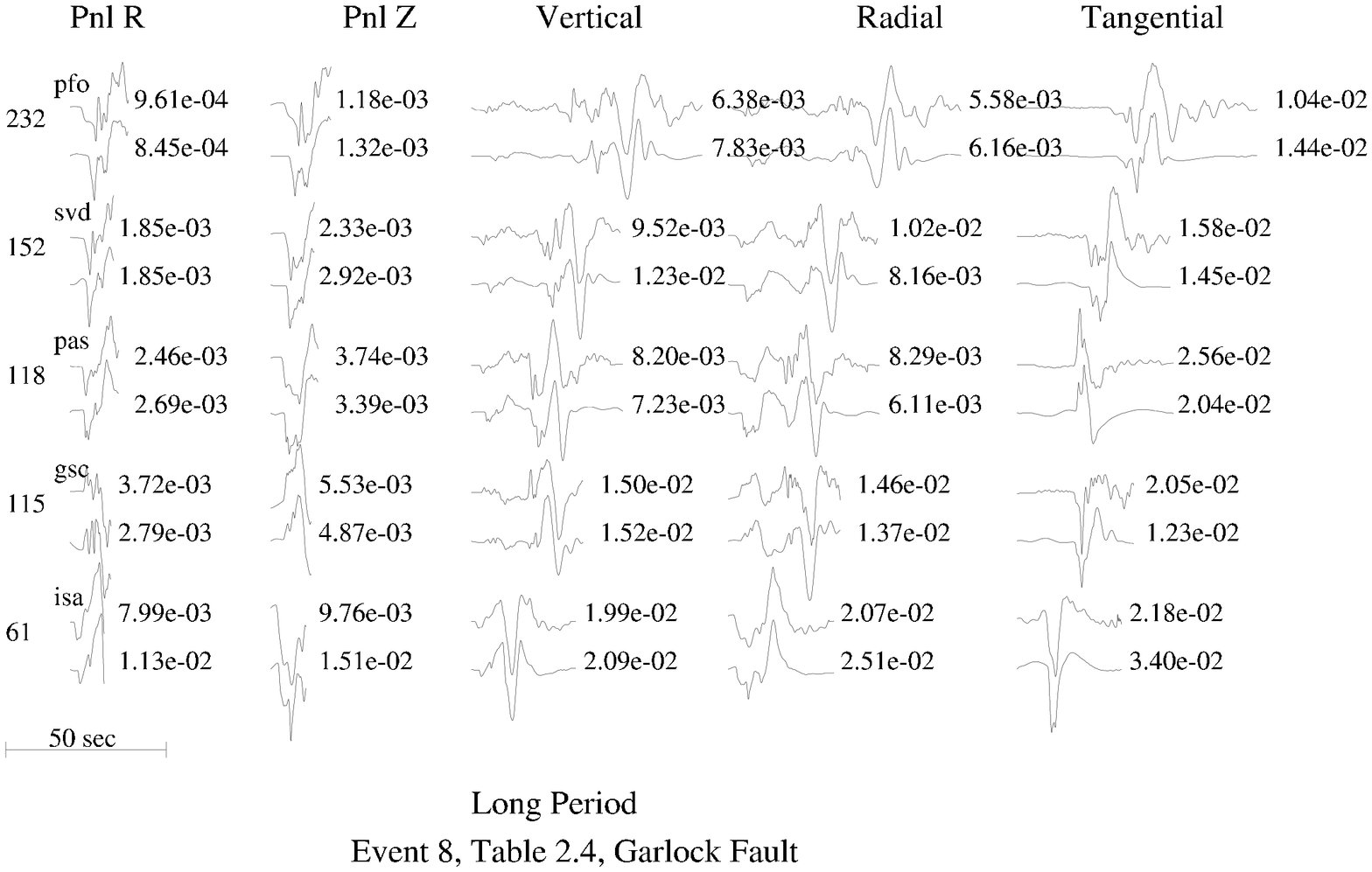,angle=270,width=6.00in}}}
\caption{
(b) Long--period waveform modeling for the July 11, 1992,
Garlock earthquake. Moment is $M_o=9.44\pm2.29\times10^{23}$
for the long--period solution.
}
\end{figure}

\clearpage


\begin{figure}
\centerline{\vbox{
\psfig{figure=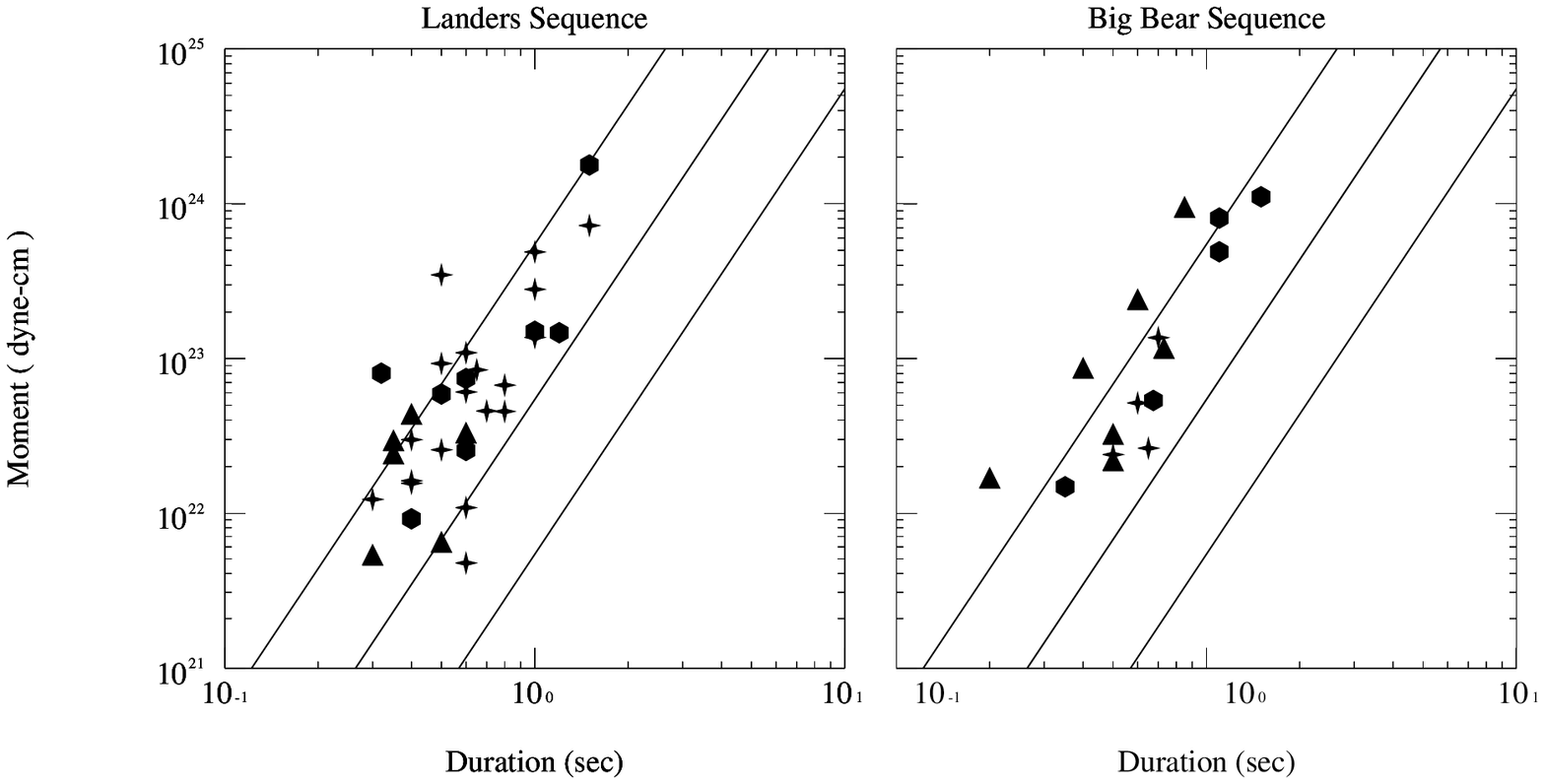,angle=270,width=6.00in}}}
\caption{
Moments versus durations for Big Bear and Landers aftershocks.
Event depths are indicated as follows: filled triangles indicate
deep events (12 to 17 km); filled crosses are intermediate (8 to 11 km)
and filled hexagons are shallow (2 to 7 km). Lines of
constant stress drop are plotted diagonally; from
bottom to top: 1, 10, 100 bars.  Figure after Jones and
Helmberger, 1996.
}
\end{figure}
\end{document}